\documentclass[11pt,a4paper]{article}
\usepackage{amssymb,amsmath,amsthm}
\usepackage[dvips]{graphicx,color}
\newcommand{\mc}{\mathcal}

\newcommand{\Uns}{\Upsilon_{\text{NS}}}
\newcommand{\Ur}{\Upsilon_{\text{R}}}

\title{\bf\LARGE A common limit of super Liouville theory and minimal models \bigskip}
\date{June 2007}
\author{Stefan Fredenhagen$^1$ and David Wellig$^2$ \\[10mm]
$^1$ Max-Planck-Institut f{\"u}r Gravitationsphysik, Albert-Einstein-Institut\\
D--14424 Golm, Germany\\[2mm]
$^{2}$ Institut f{\"u}r Theoretische Physik, ETH Z{\"u}rich\\
CH--8093 Z{\"u}rich, Switzerland}

\begin{document}
\begin{titlepage}
\maketitle
\thispagestyle{empty}

\vskip1cm

\begin{abstract}
We show that $N=1$ supersymmetric Liouville theory can be continued to
central charge $c=3/2$, and that the limiting non-rational
superconformal field theory can also be obtained as a limit of
supersymmetric minimal models. This generalises a result known for the
non-supersymmetric case. We present explicit expressions for the
three-point functions of bulk fields, as well as a set of
superconformal boundary states. The main technical ingredient to take
the limit of minimal models consists in determining analytic
expressions for the structure constants. In the appendix we show in
detail how the structure constants of supersymmetric and Virasoro
minimal models can be rewritten in terms of Barnes' double
gamma functions. 
\end{abstract}

\vspace*{-16.5cm}\noindent 
{\hfill \tt {AEI-2007-039}} \\
\bigskip\vfill
\noindent

\end{titlepage} 

\tableofcontents

\section{Introduction}

The classification of conformal theories for a given central charge is
a hard and in general maybe unmanageable task. For unitary theories, the
classification has only been achieved for central charge $c<1$ resulting
in the well known ADE series of minimal
models~\cite{Cappelli:1987xt,Kato:1987td}. One might hope that the
problem still remains tractable for the limiting value
$c=1$. In~\cite{Ginsparg:1987eb} the moduli space of $c=1$ models
based on the free boson was presented. At some points of the moduli
space, the theory is rational, and it was argued
in~\cite{Kiritsis:1988et} that these are already all rational theories
at $c=1$. That there are further non-rational models that cannot be
obtained from the free boson was shown by Runkel and
Watts~\cite{Runkel:2001ng} by explicitly constructing a new not even
quasi-rational theory at $c=1$.

Runkel and Watts considered the
unitary minimal models and defined a new theory as the limit of
minimal models when the central charge approaches $c=1$.  They
provided an explicit expression for the structure constants, and gave
strong evidence that the resulting non-rational theory is
crossing-symmetric. Later in~\cite{Schomerus:2003vv}, Schomerus
considered a continuation of Liouville theory to central charge $c=1$,
and found that it agrees with the Runkel-Watts
theory. In~\cite{Runkel:2001ng}, the authors also constructed boundary
states for this theory as a limit of minimal model boundary states
(see also~\cite{Graham:2001tg}). It was then understood
in~\cite{Fredenhagen:2004cj} that these boundary states can be
obtained from the ZZ boundary states~\cite{Zamolodchikov:2001ah} in
Liouville theory. There, also the limit of the FZZT boundary
states~\cite{Fateev:2000ik,Teschner:2000md} was constructed and it was
shown how these are obtained from the minimal model side.
\medskip

In this paper we shall construct a similar limit of minimal models in
the $N=1$ supersymmetric case. The result is a non-rational theory at
the critical central charge $c=\frac{3}{2}$ (below this value all unitary
theories fall into the rational minimal model
series~\cite{Cappelli:1986ed}). It is the first example of a unitary
superconformal theory at $c=\frac{3}{2}$ that is not part of the
moduli space of supersymmetric $c=\frac{3}{2}$ theories that are
obtained from a free boson and a free fermion~\cite{Dixon:1988ac}.
As in the bosonic case, the limiting theory can also
be obtained from $N=1$ supersymmetric Liouville theory. 
\medskip

The main technical ingredient in the computation of Runkel and Watts
was the unpublished observation of Dotsenko that the structure
constants of minimal models can be written as analytic functions
depending on a certain combination of the field labels. In appendix~A
we present a derivation of this fact. This then allows the
continuation of the structure constants to the limiting theory.

In the $N=1$ supersymmetric case, the minimal model structure
constants (in the Neveu-Schwarz sector) have been determined
in~\cite{Kitazawa:1987za} (see also~\cite{Alvarez-Gaume:1991bj}). We
derive here an analytic expression for these structure constants by
rewriting them in terms of Barnes' double gamma functions, special
functions that naturally appear in Liouville theory. The limit can
then be taken in close analogy to the derivation
in~\cite{Runkel:2001ng}. We shall provide the bulk structure constants
in the Neveu-Schwarz sector, as well as the bulk one-point functions
for a discrete family of boundary conditions.

The $N=1$ Liouville three-point functions have been computed
in~\cite{Poghosian:1996dw,Rashkov:1996jx} (see
also~\cite{Fukuda:2002bv,Belavin:2007gz}). Barnes' double gamma
functions from which they are built become singular in the limit
$c\to\frac{3}{2}$. To deal with this singularity, we make use of
asymptotic expressions for Barnes' double gamma functions that have
been derived in~\cite{Schomerus:2003vv,Fredenhagen:2004cj}. This
allows us to write the three-point function as a product of a
non-singular analytic factor, and a factor that becomes a
discontinuous but finite step function, which implements fusion
rules. The result coincides with the expressions we obtain from the
minimal models. For a discrete family of boundary conditions we
evaluate the bulk one-point function and show that it agrees with the
minimal model limit.  
\medskip

The paper is organised as follows. In section 2 we determine the limit
of supersymmetric minimal models, in section 3 the limit of super
Liouville theory is computed and compared to the minimal model
result. Section 4 summarises our results and discusses the open
problems and future directions.  Appendix A contains the derivation of
the main structural results, namely the rewriting of the (super)
minimal model structure constants in terms of Barnes' double gamma
functions. Appendix B collects some information on the special
functions that are frequently used in the paper. The last appendix~C
explicitly shows how the fusion rules pop up from the limit of
Liouville theory.

\section{Limit of super minimal models}

\subsection{Preliminaries}
The unitary, $N=1$ supersymmetric minimal models can be labelled by an
integer $p\geq 3$, the central charge is given by~\cite{Friedan:1983xq}
\begin{equation}
c = \frac{3}{2} \Big(1-\frac{8}{p (p+2)} \Big) \ .
\end{equation}
In each model there are a finite number of primary fields $\phi_{rs}$,
which are parameterised by two integers $(r,s)$ with $1\leq r\leq p-1$
and $1\leq s\leq p+1$. The combinations $(r,s)$ and $(p-r,p+2-s)$
label the same field. The conformal weight of a primary field
$\phi_{rs}$ is
\begin{equation}
h_{rs} = \frac{((p+2)r-ps)^{2}-4}{8p (p+2)} +\frac{1}{32} (1-
(-1)^{r-s}) \ .
\end{equation}
The fields $\phi_{rs}$ with $r-s$ even belong to the Neveu-Schwarz
(NS) sector, the fields with $r-s$ odd to the Ramond (R) sector. In
the Neveu-Schwarz sector, to each superconformal primary field
$\phi_{rs}$, there is a superdescendant field $\tilde{\phi}_{rs}$,
which is primary with respect to the bosonic subalgebra. Generically,
it has conformal weight $\tilde{h}_{rs}=h_{rs}+\frac{1}{2}$ (the
exception being the vacuum whose superdescendant is the supercurrent
at conformal weight $\frac{3}{2}$).

\subsection{Spectrum}
We are interested in the limit of large parameter $p$ when the central
charge approaches $c=\frac{3}{2}$. To understand how the conformal
weights of the primary fields behave in that limit, it is instructive
to rewrite the conformal weight as
\begin{align}
h^{\text{NS}}_{rs} & = \frac{d_{rs}^{2}-d_{11}^{2}}{8t} \\
h^{\text{R}}_{rs} & = \frac{d_{rs}^{2}}{8t} + \frac{c}{24}\ ,
\end{align}
with $d_{rs}=r-st$ and $t=\frac{p}{p+2}$. Because of field
identifications, we can restrict attention to labels $(r,s)$ satisfying
$r\geq st$ (where $r=st$ can only be satisfied for the Ramond ground
state, which exists for even $p$). Let us rewrite
\begin{equation}
d_{rs} = (r-s) + \frac{2}{p+2}s \ ,
\end{equation}
where the first term is an even (odd) integer in the Neveu-Schwarz (Ramond)
sector, and the second term ranges between $0$ and $2$. For a given $d$ we
look for $d_{rs}$ that approximate $d$ within a small interval of size
$\epsilon$, so we introduce the set
\begin{equation}
N (d,\epsilon ) = \big\{ (r,s) \big| d\leq d_{rs} < d + \epsilon
\big\}\ .
\end{equation}
In the following we shall restrict the discussion to the Neveu-Schwarz sector.
For $\epsilon$ small enough ($d$ and $d+\epsilon$ should have the
same even integer part) this set is given by
\begin{equation}
N (d,\epsilon) = \big\{ (2\lfloor \tfrac{d}{2}\rfloor +n,n) \big|
\{\tfrac{d}{2} \} (p+2) \leq n < (\{\tfrac{d}{2} \}+\tfrac{\epsilon}{2} ) (p+2)
\big\} \ .
\end{equation}
Here, $\lfloor x \rfloor$ is the largest integer smaller or equal to
$x$, and $\{x \}$ denotes the fractional part of $x$, $x=\lfloor
x\rfloor +\{x \}$.
For large $p$ the number of pairs $(r,s)$ contributing to $N
(d,\epsilon )$ grows as
\begin{equation}
|N (d,\epsilon)| \sim \frac{1}{2}\epsilon (p+2) \ .
\end{equation}
In particular we see that any value of $d$ can be approximated by the
$d_{rs}$ with uniform density, i.e.\ the number of
$d_{rs}$ per unit interval is independent of $d$, 
\begin{equation}\label{}
\frac{1}{\epsilon} |N (d,\epsilon)| \sim \frac{p+2}{2}\ .
\end{equation}
Following the prescription of \cite{Runkel:2001ng}, we define fields
$\phi_{d}$ of the theory at $c=\frac{3}{2}$ by averaging over fields
that lead to the same conformal weight in the limit,\footnote{for a
discussion of some aspects of limits of superminimal models where
instead of $d$ the labels $(r,s)$ are fixed
see~\cite{Rasmussen:2004gx}}
\begin{equation}\label{fielddef}
\phi_{d} (z,\bar{z}) := \lim_{\epsilon \to 0} \lim_{p\to \infty} 
\frac{n (p)}{|N (d,\epsilon)|} \sum_{(rs)\in N (d,\epsilon)} \phi_{rs}
(z,\bar{z})  \ .
\end{equation}
The function $n (p)$ will be chosen in the next subsection to get
fields in the appropriate normalisation. Note that in the following we
shall suppress the argument $\bar{z}$ in the fields~$\phi$.

\subsection{Operator product expansion} 
We require that in the limit the operator product expansion (OPE) of two
fields $\phi_{d_{i}}$ remains finite, i.e.\ it should be of the form
\begin{align}
\phi_{d_{1}} (z_{1}) \phi_{d_{2}} (z_{2}) & = \int \text{d}d_{3}
\frac{1}{|z_{12}|^{2 (h_{1}+h_{2}-h_{3})}} D (d_{1},d_{2},d_{3})
\big( \phi_{d_{3}} (z_{2}) + \dotsb \big) 
+ \text{superdescendants} \ ,
\end{align}
with some function (or distribution) $D (d_{1},d_{2},d_{3})$. The dots
stand for contributions of Virasoro descendant fields, and
$z_{12}=z_{1}-z_{2}$. From the definition~\eqref{fielddef} of the
fields $\phi_{d}$ we get
\begin{align}
\phi_{d_{1}} (z_{1}) \phi_{d_{2}} (z_{2}) &  =
\lim_{\epsilon \to 0}\lim_{p\to \infty} \frac{4n (p)^{2}}{\epsilon^{2}p^{2}}
\sum_{(r_{i}s_{i})\in N (d_{i},\epsilon)} 
\phi_{r_{1}s_{1}} (z_{1}) \phi_{r_{2}s_{2}}
(z_{2}) + \dotsb \nonumber \\
& = \lim_{\epsilon \to 0}\lim_{p\to \infty} \frac{4n (p)^{2}}{\epsilon
^{2}p^{2}} 
\sum_{(r_{i}s_{i})\in N
(d_{i},\epsilon)} \sum_{r_{3},s_{3}}
\mc{N}^{(p)}_{r_{1}r_{2}}{}^{r_{3}}
\mc{N}^{(p+2)}_{s_{1}s_{2}}{}^{s_{3}} \nonumber\\
& \quad\   \times \bigg( \frac{\delta_{2}
(k+l)}{|z_{12}|^{2 (h_{r_{1}s_{1}}+h_{r_{2}s_{2}}-h_{r_{3}s_{3}})}}
D^{\text{NS}}_{(r_{1}s_{1}) (r_{2}s_{2})}{}^{(r_{3}s_{3})}
\phi_{r_{3}s_{3}} (z_{2}) \nonumber\\
& \qquad\quad \   +\frac{\delta_{2}
(k+l+1)}{|z_{12}|^{2 (h_{r_{1}s_{1}}+h_{r_{2}s_{2}}-\tilde{h}_{r_{3}s_{3}})}}
\tilde{D}^{\text{NS}}_{(r_{1}s_{1}) (r_{2}s_{2})}{}^{(r_{3}s_{3})}
\tilde{\phi}_{r_{3}s_{3}} (z_{2}) \bigg) + \dotsb \ .  
\end{align} 
We used here the operator product expansion~\eqref{N1OPE} for minimal
model fields, which is stated in appendix~A. In the OPE the fusion
rules enter which can be expressed in terms of the fusion rules
$\mathcal{N}^{(p)}$ (given in~\eqref{fusion}) of the $su (2)$ WZW
model. $\delta_{2} (n)$ is defined to be $1$ for $n$ even, and $0$ for
$n$ odd.

The important insight is that the structure constants only depend on
the combinations $d_{r_{i}s_{i}}=r_{i}-ts_{i}$, and that this
dependence is analytic. This result is derived in appendix~A, the
final expressions for the structure constants are given
in~\eqref{mmstruc} and~\eqref{mmstrucdesc}. It can easily be seen that
the expressions have well-defined limits
$D^{\text{NS}}_{d_{1}d_{2}}{}^{d_{3}}$,
$\tilde{D}^{\text{NS}}_{d_{1}d_{2}}{}^{d_{3}}$ as $p\to \infty$,
\begin{align}\label{NSstructconst}
D^{\text{NS}}_{d_{1}d_{2}}{}^{d_{3}} & = \left[ \frac{\Upsilon
(1-\frac{\tilde{d}}{2}|1) \Upsilon (1-\frac{\tilde{d}_{1}}{2}|1)
\Upsilon (1-\frac{\tilde{d}_{2}}{2}) \Upsilon
(1-\frac{\tilde{d}_{3}}{2}|1)}{\Upsilon (1|1) \Upsilon
(1-\frac{d_{1}}{2}|1) \Upsilon (1-\frac{d_{2}}{2}|1) \Upsilon
(1-\frac{d_{3}}{2}|1)}\right]^{2} \ ,\\
\intertext{and} 
\label{NSstructconsttilde}
\tilde{D}^{\text{NS}}_{d_{1}d_{2}}{}^{d_{3}} & = \frac{1}{h_{3}}
\frac{\Upsilon (\frac{1}{2}-\frac{\tilde{d}}{2}|1) \Upsilon
(\frac{3}{2}-\frac{\tilde{d}}{2}|1) \prod_{i}\Upsilon
(\frac{1}{2}-\frac{\tilde{d}_{i}}{2}|1) \Upsilon
(\frac{3}{2}-\frac{\tilde{d}_{i}}{2}|1)}{\big[ \Upsilon (1|1) \Upsilon
(1-\frac{d_{1}}{2}|1) \Upsilon (1-\frac{d_{2}}{2}|1) \Upsilon
(1-\frac{d_{3}}{2}|1)\big]^{2}} \ .
\end{align}
Here, $2\tilde{d}=d_{1}+d_{2}+d_{3}$ and
$\tilde{d}_{i}=\tilde{d}-d_{i}$. The functions $\Upsilon (x|b)$ are
special combinations of Barnes' double gamma functions (see appendix~B).

The only thing left to consider are
the fusion rules. For given $(r_{i},s_{i})\in N (d_{i},\epsilon)$,
$i=1,2$, we shall analyse which labels $(r_{3},s_{3})$ appear in the
sum above due to the fusion rules $\mc{N}^{(p)}\mc{N}^{(p+2)}$. The
label $(r_{3},s_{3})$ appears if (up to corrections of order
$\epsilon$ or $1/p$)
\begin{equation}\label{fusionlimit}
| \{\tfrac{d_{1}}{2} \} - \{\tfrac{d_{2}}{2} \}| < \{ \tfrac{d_{3}}{2} \}
< \min (\{\tfrac{d_{1}}{2} \}+\{\tfrac{d_{2}}{2}
\},2-\{\tfrac{d_{1}}{2}\}- \{\tfrac{d_{2}}{2} \}) \ \text{and}\ 
s_{1}+s_{2}+s_{3} \ \text{odd} \ .
\end{equation}
The condition on the $s_{i}$ leads to the conclusion that for given
$(r_{1},s_{1})$ and $(r_{2},s_{2})$ only one half of the labels in $N
(d_{3},\epsilon )$ appear if $d_{3}$ satisfies the inequality
in~\eqref{fusionlimit}. (Note that here we allow also for negative
$d_{3}$, the definition of $N (d_{3},\epsilon)$ can be extended to
this case in an obvious way.) As we sum over $(r_{1},s_{1})\in N
(d_{1},\epsilon)$ and $(r_{2},s_{2})\in N (d_{2},\epsilon)$, each
label $(r_{3},s_{3})$ in $N (d_{3},\epsilon)$ appears with
multiplicity $\frac{1}{2}|N (d_{1},\epsilon )||N (d_{2},\epsilon)|$.

The condition on $k+l$ can be reformulated in terms of the integer parts
$\lfloor \frac{d_{i}}{2}\rfloor$,
\begin{equation}
k+l = \lfloor \tfrac{d_{1}}{2}\rfloor  + \lfloor \tfrac{d_{2}}{2}
\rfloor  - \lfloor \tfrac{d_{3}}{2}\rfloor  +
s_{1}+s_{2}-s_{3} +1 \ .
\end{equation}
Using that $\sum s_{i}$ is odd~\eqref{fusionlimit}, the condition
$k+l$ even or odd, translates directly into a condition on
$\sum_{i}\lfloor \tfrac{d_{i}}{2}\rfloor$. We find thus
\begin{align}
\phi_{d_{1}} (z_{1}) \phi_{d_{2}} (z_{2})  =
\lim_{\epsilon \to 0}\lim_{p\to \infty} n (p)^{2}
\frac{1}{2} 
\sideset{}{'}\sum_{d_{3},\epsilon}
& \sum_{(r_{3},s_{3})\in N (d_{3},\epsilon)} 
\bigg( \frac{\delta_{2} (\sum \lfloor
\frac{d_{i}}{2}\rfloor)}{|z_{12}|^{2 (h_{1}+h_{2}-h_{3})}}
D^{\text{NS}}_{d_{1}d_{2}}{}^{d_{3}}
\phi_{r_{3}s_{3}} (z_{2}) \nonumber\\
& \qquad  +\frac{\delta_{2} (\sum \lfloor
\frac{d_{i}}{2}\rfloor+1)}{|z_{12}|^{2 (h_{1}+h_{2}-\tilde{h}_{3})}}
\tilde{D}^{\text{NS}}_{d_{1}d_{2}}{}^{d_{3}}
\tilde{\phi}_{r_{3}s_{3}} (z_{2}) \bigg) + \dotsb \ .
\end{align}
The primed sum indicates that we only sum over those $d_{3}$ that
satisfy the inequality in~\eqref{fusionlimit} and the $\epsilon$ indicates that
$d_{3}$ is summed over in steps of $\epsilon$. We would like to turn
this sum into an integral in the limit $\epsilon \to 0$ which amounts
to replacing the sum by $\frac{1}{\epsilon} \int \text{d}d_{3}$. The sum over
the $\phi_{r_{3}s_{3}}$ weighted by the factor $n (p)/|N|$
turns into the field $\phi_{d_{3}}$. So if we choose $n (p)$
such that
\begin{equation}
n (p) |N| \epsilon^{-1} \to 1
\end{equation}
we obtain a finite OPE in the limit. This amounts to setting 
\begin{equation}
n (p) = \frac{2}{p} \ .
\end{equation}
We finally obtain
\begin{align}\label{OPElimit}
\phi_{d_{1}} (z_{1}) \phi_{d_{2}} (z_{2}) & =
\int_{\mathbb{R}_{+}} {\rm d}d_{3} \, \tfrac{1}{2} 
\bigg( \frac{1}{|z_{12}|^{2 (h_{1}+h_{2}-h_{3})}} P
(\tfrac{d_{1}}{4},\tfrac{d_{2}}{4},\tfrac{d_{3}}{4})
D^{\text{NS}}_{d_{1}d_{2}}{}^{d_{3}}
\phi_{d_{3}} (z_{2}) \nonumber\\
&\qquad\quad  +\frac{1}{|z_{12}|^{2 (h_{1}+h_{2}-\tilde{h}_{3})}} P
(\tfrac{d_{1}+2}{4},\tfrac{d_{2}+2}{4},\tfrac{d_{3}+2}{4})
\tilde{D}^{\text{NS}}_{d_{1}d_{2}}{}^{d_{3}} \tilde{\phi}_{d_{3}}
(z_{2}) \bigg)+ \dotsb \ .
\end{align}
Here we restricted the integration domain to the positive numbers
using the field identification $d_{3}\to -d_{3}$. The structure
constants $D^{\text{NS}}_{d_{1}d_{2}}{}^{d_{3}}$ and
$\tilde{D}^{\text{NS}}_{d_{1}d_{2}}{}^{d_{3}}$ in the limit $p\to
\infty$ (given in~\eqref{NSstructconst}
and~\eqref{NSstructconsttilde}) are invariant under the replacement
$d_{3}\to -d_{3}$ (which follows from $\Upsilon (x|1)=\Upsilon
(2-x|1)$, see appendix~B). The function $P$ implements the fusion
rules; it is a step function taking the values $0$ and $1$, its
definition is given in~\eqref{defofP}.

\subsection{Two-point functions}
Let us now discuss the correlation functions. We want a normalisation
such that
\begin{equation}\label{twopointfunction}
\langle \phi_{d_{1}} (z_{1}) \phi_{d_{2}}
(z_{2})\rangle = \frac{\delta
(d_{1}-d_{2})}{|z_{12}|^{4h_{1}}} \ .
\end{equation}
To obtain this we have to rescale the correlators, which corresponds to a
change of normalisation of the bulk vacuum. Let us denote the scaling
factor for the correlation functions by $\gamma^{2} (p)$. Then the two-point
correlator in the limiting theory is defined as
\begin{align}
\langle \phi_{d_{1}} (z_{1})\phi_{d_{2}} (z_{2})\rangle
:= &\ \lim_{\epsilon \to 0} \lim_{p\to \infty} \frac{\gamma^{2} (p) 
n(p)^{2}}{|N |^{2}} \sum_{(r_{i}s_{i})\in N(d_{i},\epsilon)} \langle
\phi_{r_{1}s_{1}} (z_{1}) \phi_{r_{2}s_{2}}
(z_{2})\rangle \nonumber\\
= & \ \lim_{\epsilon \to 0} \lim_{p\to \infty} \frac{16 \gamma^{2}
(p)}{\epsilon^{2}p^{4}}   
\sum_{(r_{1}s_{1})\in N(d_{1},\epsilon)\cap N(d_{2},\epsilon)}
\frac{1}{|z_{12}|^{4h_{1}}} \ , \label{twopoint}
\end{align}
where we used that the minimal model fields have normalised two-point
functions, 
\begin{equation}\label{minmodtwopoint}
\langle \phi_{r_{1}s_{1}} (z_{1})\phi_{r_{2}s_{2}}
(z_{2})\rangle =
\frac{\delta_{r_{1}r_{2}}\delta_{s_{1}s_{2}}}{|z_{12}|^{4h_{1}}} \ .
\end{equation}
(This is valid if both $r_{i}>s_{i}t$, otherwise one has to take
field identification into account.) For $\epsilon$ small enough, the
number of elements in the intersection of the sets 
$N(d_{i},\epsilon)$ is given by
\begin{equation}\label{intersection}
|N (d_{1},\epsilon) \cap N (d_{2},\epsilon)| \sim \frac{p+2}{2}
(\epsilon -|d_{1}-d_{2}|) \Theta (\epsilon -|d_{1}-d_{2}|) \ ,
\end{equation}
where $\Theta (x)=1$ for $x\geq 0$ and $0$ otherwise. The factor of
$p/2$ in~\eqref{intersection} together with the $16 p^{-4}$ is absorbed
by choosing 
\begin{equation}\label{defofgamma}
\gamma^{2}(p)= (p/2)^{3} \ .
\end{equation}
The remaining function $\epsilon^{-2} (\epsilon -|x|)\Theta (\epsilon
-|x|)$ converges to the delta distribution $\delta (x)$, so that we
obtain the desired normalisation~\eqref{twopointfunction}.

\subsection{Three-point functions} 

With the operator product expansion in~\eqref{OPElimit} and the
normalisation of the two-point function, we can directly write down
the three-point correlation functions. For three superconformal
primary fields, we find
\begin{align}
\langle \phi_{d_{1}} (z_{1}) \phi_{d_{2}}
(z_{2})\phi_{d_{3}} (z_{3}) \rangle
& = \frac{1}{2} P (\tfrac{d_{1}}{4},\tfrac{d_{2}}{4},\tfrac{d_{3}}{4})
\, D^{\text{NS}}_{d_{1}d_{2}}{}^{d_{3}}
\, |z_{12}|^{(d_{3}^{2}-d_{1}^{2}-d_{2}^{2})/4} \nonumber\\
& \quad\ \times  
|z_{13}|^{(d_{2}^{2}-d_{1}^{2}-d_{3}^{2})/4}
|z_{23}|^{(d_{1}^{2}-d_{2}^{2}-d_{3}^{2})/4}\ .
\label{threepoint}
\end{align}
For two primary fields and one superdescendant field we obtain
similarly 
\begin{align}
\langle \phi_{d_{1}} (z_{1}) \phi_{d_{2}} (z_{2})\tilde{\phi}_{d_{3}}
(z_{3})\rangle & = \frac{1}{2} P
(\tfrac{d_{1}+2}{4},\tfrac{d_{2}+2}{4},\tfrac{d_{3}+2}{4})
\, \tilde{D}^{\text{NS}}_{d_{1}d_{2}}{}^{d_{3}} \, 
|z_{12}|^{(d_{3}^{2}+4-d_{1}^{2}-d_{2}^{2})/4}\nonumber\\
& \quad\  \times 
|z_{13}|^{(d_{2}^{2}-d_{1}^{2}-d_{3}^{2}-4)/4}
|z_{23}|^{(d_{1}^{2}-d_{2}^{2}-d_{3}^{2}-4)/4}\ .
\label{threepointdesc}
\end{align}

\subsection{One-point functions on the upper half plane}

Boundary conditions in supersymmetric minimal models that preserve the
superconformal symmetry have been analysed
in~\cite{Nepomechie:2001bu} (for $p$ odd). The boundary conditions are
parameterised by the Kac labels. For Kac labels $(u,v)$ with $u+v$ even
(NS sector), we have two boundary theories $(u,v)_{\pm}$ corresponding
to brane and anti-brane. For these boundary conditions the one-point
functions for NSNS bulk fields $\phi_{rs}$ ($r+s$ even) are given
by\footnote{To distinguish NSNS and RR field we shall use superscripts
$^{\text{NS}}$ and $^{\text{R}}$ frequently in this subsection.} 
\begin{equation}
\langle \phi^{\text{NS}}_{rs} (z) \rangle_{(u,v)_{\pm}} =
\frac{1}{|z-\bar{z}|^{2h_{rs}}} \frac{\sqrt{2}}{\sqrt[4]{p (p+2)}}
\frac{\sin \frac{\pi ru}{p}\sin \frac{\pi sv}{p+2}}{\sqrt{\sin
\frac{\pi r}{p}\sin \frac{\pi s}{p+2}}} \ ,
\end{equation}
for RR fields ($r+s$ odd) they read
\begin{equation}\label{RRone-point}
\langle \phi^{\text{R}}_{rs} (z) \rangle_{(u,v)_{\pm}} = \pm 
\frac{1}{|z-\bar{z}|^{2h_{rs}}} \frac{2^{\frac{5}{4}}}{\sqrt[4]{p (p+2)}}
(-1)^{\frac{u-v}{2}}
\frac{\sin \frac{\pi ru}{p}\sin \frac{\pi sv}{p+2}}{\sqrt{\sin
\frac{\pi r}{p}\sin \frac{\pi s}{p+2}}} \ .
\end{equation}
The normalisation of the RR fields is chosen such they have the
standard two-point function~\eqref{minmodtwopoint}.
There are also boundary theories associated to Ramond labels $(u,v)$
with $u+v$ odd. They correspond to the opposite sign in the gluing
condition for the supercurrent at the boundary. We shall not consider
those here, but concentrate in the following on the boundary theories
$(u,v)_{\pm}$ with $u+v$ even.

When we now discuss boundary conditions in the limit $p\to \infty$, we
essentially have two options. We can take the boundary labels $(u,v)$
fixed, or we can scale them such that the conformal weight
$h_{uv}$ corresponding to the Kac labels stays fixed in the limit. For bosonic
minimal models, both options have been investigated leading to one
discrete family of boundary
conditions~\cite{Graham:2001tg,Runkel:2001ng}, and to one continuous
family~\cite{Fredenhagen:2004cj}. For the supersymmetric minimal
models we shall only consider the case when the boundary labels are
taken to be fixed.

To analyse the limit $p\to\infty$ we should try to rewrite the one-point
function for $\phi_{rs}$ such that it depends on $d_{rs}=r-st$. This
is indeed possible due to the identity (for $u+v$ even)
\begin{equation}\label{trigidentity}
\sin \frac{\pi ur}{p} \sin \frac{\pi vs}{p+2} = \left\{\begin{array}{ll}
\sin \tfrac{\pi ud_{rs}}{2t} \sin \tfrac{\pi vd_{rs}}{2} & \text{for}\
r+s\ \text{even}\\[1mm]
(-1)^{\frac{u-v}{2}} \sin\big( \tfrac{\pi ud_{rs}}{2t} +\tfrac{\pi
uv}{2} \big) \sin \big(\tfrac{\pi v d_{rs}}{2} +\tfrac{\pi uv}{2}
\big) & \text{for}\ r+s \ \text{odd} \ .
\end{array} \right.
\end{equation}
Let us now consider the one-point function for a NSNS bulk field
$\phi^{\text{NS}}_{d}$ in the presence of a boundary condition
$(u,v)_{\pm}$. Using the definition~\eqref{fielddef} of
$\phi^{\text{NS}}_{d} (z)$ and suppressing the obvious $z$-dependence
we obtain
\begin{align}
\langle \phi^{\text{NS}}_{d} \rangle_{(u,v)_{\pm}} & =
\lim_{p\to\infty} n (p) \gamma
(p) \sqrt{\frac{2}{p}} \, \frac{\sin \frac{\pi u d}{2} \sin \frac{\pi
vd}{2}}{|\sin \frac{\pi d}{2}|} \nonumber \\
& = \frac{\sin \frac{\pi u d}{2} \sin \frac{\pi
vd}{2}}{|\sin \frac{\pi d}{2}|} \ .
\label{NSone-point-limit}
\end{align}
Note that we had to rescale the disc correlator by the factor
$\gamma(p)$ (given in~\eqref{defofgamma}) that we introduced to
obtain correctly normalised sphere correlators. The factor $\gamma
(p)$ can be thought of as rescaling the bulk vacuum such that
sphere correlators obtain a factor $\gamma^{2} (p)$ while disc
correlators are changed by $\gamma (p)$. (This can be understood best
from the boundary state formalism, see section~4.2
of~\cite{Runkel:2001ng} for a discussion.) 

The one-point functions obtained above are not independent. Because of
the identity
\begin{equation}
\sin \pi u x \, \sin \pi v x = \sum_{\ell =|u-v|+1,2}^{u+v-1} \sin \pi
\ell x \, \sin \pi x \ ,
\end{equation}
where the summation variable $\ell$ is increased in steps of $2$, we can
express every one-point function as a sum of one-point functions for
boundary labels $(\ell ,1)_{\pm}$,
\begin{equation}
\langle \phi^{\text{NS}}_{d}\rangle_{(u,v)_{\pm}} 
= \sum_{\ell =|u-v|+1,2}^{u+v-1} 
\langle \phi^{\text{NS}}_{d}\rangle_{(\ell ,1)_{\pm}} \ .
\end{equation} 
This suggests that only the boundary conditions $(\ell,1)_{\pm}$ are
elementary. We have to check that a similar condition is true for the
one-point functions of RR fields. 

To obtain the one-point functions for RR fields in the limit
$p\to\infty$ we should again rewrite the one-point
function~\eqref{RRone-point} for $\phi^{\text{R}}_{rs}$ in terms of
$d_{rs}$. This is possible due to the
identity~\eqref{trigidentity}. Using the same definition (including
the normalisation) of $\phi^{\text{R}}_{d}$ in terms of
$\phi^{\text{R}}_{rs}$ as for the NS case (see~\eqref{fielddef}) we
find
\begin{equation}\label{RRone-point-limit}
\langle \phi^{\text{R}}_{d}\rangle_{(u,v)_{\pm}}= \pm
2^{\frac{3}{4}}\frac{\sin \frac{\pi}{2} (u d +uv)\sin \frac{\pi}{2}
(vd+uv)}{|\cos \frac{\pi}{2}d|} \ .
\end{equation}
Again we find that the one-point functions are not independent. For
$u,v$ odd, the numerator is most easily expressed by cosine-functions,
and we have
\begin{equation}
\cos \pi u x \, \cos \pi v x = \sum_{\ell =|u-v|+1,2}^{u+v-1}
(-1)^{\frac{\ell -|u-v|-1}{2}} \cos \pi \ell x \, \cos \pi x \ .
\end{equation}
For $u,v$ even we find
\begin{equation}
\sin \pi ux \, \sin \pi vx = \sum_{\ell =|u-v|+1,2}^{u+v-1}
(-1)^{\frac{\ell -|u-v|-1}{2}} \cos \pi \ell x \, \cos \pi x \ .
\end{equation}
This shows that the boundary conditions
$(u,v)_{\pm}$ really are superpositions of boundary conditions $(\ell
,1)_{\pm}$, more precisely we have
\begin{equation}
(u,v)_{+} = (|u-v|+1,1)_{+} \oplus (|u-v|+3,1)_{-} \oplus \dotsb
\oplus (u+v-1,1)_{(-)^{uv}} \ .
\end{equation}

\section{Limit of super Liouville theory}

\subsection{Preliminaries}
The supersymmetric extension of Liouville theory was first considered
in~\cite{Polyakov:1981re}. Shortly after the discovery of the exact
three-point function of bosonic Liouville theory
in~\cite{Dorn:1994xn,Zamolodchikov:1995aa}, the exact bulk structure
constants for the supersymmetric version were found
in~\cite{Poghosian:1996dw,Rashkov:1996jx}. The theory depends on a
parameter $b$ that determines the central charge,
\begin{equation}
c=\frac{3}{2} (1+2Q^{2}) \quad \text{with}\  Q=b+\frac{1}{b} \ .
\end{equation}
The primary fields $V^{\text{NS}}_{\alpha}$ in the NSNS sector are
labelled by a parameter $\alpha=\frac{Q}{2}+ip$ with real momentum
$p$. The conformal weight of~$V^{\text{NS}}_{\alpha}$ is given by
\begin{equation}
h_{\alpha} = \frac{\alpha (Q-\alpha)}{2} = \frac{Q^{2}}{8}
+\frac{p^{2}}{2}\ .
\end{equation}
The superdescendant fields $\tilde{V}^{\text{NS}}_{\alpha}$ then have
conformal weight $h_{\alpha}+\frac{1}{2}$.

In the Ramond-Ramond sector, the ground state of a representation 
generically has degeneracy $4$ due to the presence of zero modes of the
supercurrent. The corresponding RR fields are labelled by
$\Theta^{\epsilon ,\bar{\epsilon}}_{\alpha}$ with $\epsilon
,\bar{\epsilon}=\pm 1$; the OPE with the supercurrent reads
\begin{align}
T_{\text{F}} (z) \Theta_{\alpha}^{\epsilon ,\bar{\epsilon}} (0) & \sim
\frac{p\, \Theta_{\alpha}^{-\epsilon ,\bar{\epsilon}}}{\sqrt{2}z^{3/2}} & 
-i\Theta_{\alpha}^{\epsilon ,\bar{\epsilon}} (0) \bar{T}_{\text{F}}
(\bar{z}) & \sim \frac{p\, \Theta_{\alpha}^{\epsilon
,-\bar{\epsilon}}}{\sqrt{2}\bar{z}^{3/2}} \ . 
\end{align}
Similar to the theory of a free fermion (see e.g.\
\cite{Ginsparg:1988ui}), locality and modular invariance restrict
these four fields to one combination $V^{\text{R}}_{\alpha }$, which can
be chosen as
\begin{equation}\label{newRRfields}
V^{\text{R}}_{\alpha} = \frac{1}{\sqrt{2}}\big( \Theta_{\alpha}^{++} +
\Theta_{\alpha}^{--} \big) \ .
\end{equation}
The conformal weight of $V^{\text{R}}_{\alpha}$ is
\begin{equation}
h^{\text{R}}_{\alpha} = h_{\alpha} +\frac{1}{16} = \frac{p^{2}}{2} +
\frac{c}{24} \ .
\end{equation}

\subsection{Two- and three-point functions: Neveu-Schwarz sector}
The three-point function of primary fields in the Neveu-Schwarz sector
reads~\cite{Fukuda:2002bv}
\begin{equation}
\langle V^{\text{NS}}_{\alpha_{1}}(z_{1})V^{\text{NS}}_{\alpha_{2}}
(z_{2})V^{\text{NS}}_{\alpha_{3}} (z_{3}) \rangle = C^{\text{NS}}
(\alpha_{i}) |z_{12}|^{2 (h_{1}+h_{2}-h_{3})}
|z_{23}|^{2 (h_{2}+h_{3}-h_{1})} |z_{13}|^{2 (h_{1}+h_{3}-h_{2})}\ ,
\end{equation}
with
\begin{equation}
C^{\text{NS}} (\alpha_{i}) = \big(\mu \pi \gamma (\tfrac{bQ}{2}) b^{1-b^{2}}
\big)^{\frac{Q-2\tilde{\alpha}}{b}} \, \frac{\Uns ' (0) \Uns
(2\alpha_{1})\Uns (2\alpha_{2})\Uns (2\alpha_{3})}{\Uns
(2\tilde{\alpha}-Q) \Uns (2\tilde{\alpha}_{1}) \Uns
(2\tilde{\alpha}_{2}) \Uns (2\tilde{\alpha}_{3})} \ .
\end{equation}
Here, $2\tilde{\alpha} =\alpha_{1}+\alpha_{2}+\alpha_{3}$ and 
$\tilde{\alpha}_{i}=\tilde{\alpha} -\alpha_{i}$. The
functions $\Uns$ are combinations of two $\Upsilon$ functions (see
appendix~B),
\begin{align}\label{defofUnsUr}
\Uns (x) & = \Upsilon \big(\tfrac{x}{2}\big| b \big) \Upsilon
\big(\tfrac{x+Q}{2}\big| b \big) & \Upsilon_{\text{R}} = \Upsilon
\big(\tfrac{x+b}{2}\big| b \big) \Upsilon
\big(\tfrac{x+b^{-1}}{2}\big|b \big) \ ,
\end{align}
where we also introduced the functions $\Upsilon_{\text{R}}$ that we
shall need later. The momenta $\alpha_{i}$ are of the form
\begin{equation}
\alpha_{i}=\frac{Q}{2} +i p_{i} \ ,
\end{equation}
with real $p_{i}$. The normalisation is such that the two-point
function is given by~\cite{Fukuda:2002bv}
\begin{equation}
\langle V^{\text{NS}}_{\alpha_{1}} (z_{1}) V^{\text{NS}}_{\alpha_{2}}
(z_{2})\rangle = 
|z_{12}|^{4h_{\alpha_{1}}} 2\pi \big(\delta (Q-\alpha_{1}-\alpha_{2}) +
\delta (\alpha_{1}-\alpha_{2}) R^{\text{NS}} (\alpha_{1}) \big) \ ,
\end{equation}
with the reflection amplitude
\begin{equation}
R^{\text{NS}} (\alpha) = - \big( \mu \pi \gamma
(\tfrac{bQ}{2})\big)^{\frac{Q-2\alpha}{b}} \frac{\Gamma (b (\alpha
-\frac{Q}{2})) \Gamma (\frac{1}{b} (\alpha -\frac{Q}{2}))}{\Gamma (-b
(\alpha -\frac{Q}{2})) \Gamma (-\frac{1}{b} (\alpha -\frac{Q}{2}))} \ .
\end{equation}
To get to a theory with central charge $c=\frac{3}{2}$, we have to take
the limit $b\to i$. To perform this limit we need the asymptotics of the
functions $\Upsilon$ from~\cite{Fredenhagen:2004cj},
\begin{equation}\label{asympUpsilon}
\Upsilon (ip +\kappa Q +\mathcal{O} (Q^{2})) = \frac{1}{2}
e^{-\frac{1}{\epsilon} (\lambda (p)-\frac{\pi^{2}}{2})} e^{i\pi
\frac{p^{2}}{2}} e^{-i\pi (p-2\kappa +1) (p-\lfloor
p\rfloor-\frac{1}{2})} \Upsilon (1-p|1)^{-1}\ (1+o (\epsilon^{0})) \ ,
\end{equation}
where $\lambda (p)=2\pi^{2} (p-\lfloor p\rfloor -\frac{1}{2})^{2}$,
and $\epsilon =2\pi i Q/b$. (The above asymptotics strictly hold only
for non-integer $p$, which will be enough for our purposes,
see~\cite{Fredenhagen:2004cj} for further details.) Using this we can
determine the asymptotics of the functions $\Uns$ and
$\Upsilon_{\text{R}}$,
\begin{subequations}
\label{asympU}
\begin{align}
\label{asympUns1}
\Uns (Q+2ip) & \sim \frac{1}{4} e^{-\frac{2}{\epsilon} (\lambda
(p)-\frac{\pi^{2}}{2})} e^{i\pi p^{2}} e^{i\pi (1-2p)
(p-\lfloor p\rfloor -\frac{1}{2})} \, \Upsilon (1-p|1)^{-2} \\
\label{asympUns2} 
\Uns (\tfrac{Q}{2}+2ip) & \sim \frac{1}{4}
e^{-\frac{2}{\epsilon} (\lambda (p)-\frac{\pi^{2}}{2})}
e^{i\pi p^{2}} e^{-2\pi i p (p-\lfloor
p\rfloor -\frac{1}{2})} \, \Upsilon (1-p|1)^{-2} \\
\label{asympUr1}
\Upsilon_{\text{R}} (Q+2ip) & \sim \frac{1}{4} e^{-\frac{2}{\epsilon}
(\lambda (p+\frac{1}{2})-\frac{\pi^{2}}{2})} e^{i\pi
p^{2}+\frac{i\pi}{4}} e^{i\pi (1-2p) (p-\lfloor
p+\frac{1}{2}\rfloor)} \, [\Upsilon (\tfrac{1}{2}-p|1) \Upsilon
(\tfrac{3}{2}-p|1)]^{-1}\\
\label{asympUr2}
\Ur (\tfrac{Q}{2}+2ip) & \sim \frac{1}{4} e^{-\frac{2}{\epsilon} (\lambda
(p+\frac{1}{2})-\frac{\pi^{2}}{2})} e^{i\pi
p^{2}+\frac{i\pi}{4}} e^{-2\pi ip (p-\lfloor
p+\frac{1}{2}\rfloor )} \, [\Upsilon (\tfrac{1}{2}-p|1) \Upsilon
(\tfrac{3}{2}-p|1)]^{-1} \ .
\end{align}
\end{subequations}
Moreover we need
\begin{equation}
\Upsilon ' (0|b) = \frac{2\pi^{2}}{\epsilon} \Upsilon (1|1)^{-1}
(1+o (\epsilon^{0})) \ .
\end{equation}
which leads to
\begin{equation}
\Uns ' (0) = \frac{\pi^{2}}{2\epsilon} \Upsilon (1|1)^{-2} (1+o
(\epsilon^{0})) \ .
\end{equation}
Putting everything together we can now find the asymptotics of the
structure constants,
\begin{align}
C^{\text{NS}} (\alpha_{i}) & = \big(\mu \pi \gamma (\tfrac{bQ}{2})
\big)^{\frac{Q-2\tilde{\alpha}}{b}} \, \frac{2\pi^{2}}{\epsilon}
e^{-\frac{2}{\epsilon}F (\frac{p_{i}}{2})} 
e^{i\eta (p_{i})} \nonumber\\
&\quad\ \times  \bigg[
\frac{\Upsilon (1-\tilde{p}|1)\Upsilon (1-\tilde{p}_{1}|1)\Upsilon
(1-\tilde{p}_{2}|1) \Upsilon (1-\tilde{p}_{3}|1)}{\Upsilon
(1|1)\Upsilon (1-p_{1}|1)\Upsilon (1-p_{2}|1) \Upsilon (1-p_{3}|1)}
\bigg]^{2} (1+o (\epsilon^{0})) \ .
\label{scasymptotics}
\end{align}
Here, $e^{i\eta (p_{i})}$ contains the phase factors from the
asymptotics of the functions $\Uns$ as well as the limit of
$b^{(1-b^{2})\frac{Q-2\tilde{\alpha}}{b}}$, and $F$ is given by 
\begin{equation}\label{defofF}
F (\tfrac{p_{i}}{2}) = \lambda (p_{1}) +\lambda (p_{2})+\lambda
(p_{3}) -\lambda (\tilde{p}) -\lambda (\tilde{p}_{1}) -\lambda
(\tilde{p}_{2}) -\lambda (\tilde{p}_{3}) +\tfrac{\pi^{2}}{2} \ .
\end{equation}
This function is always non-negative, so the limit
$P (\frac{p_{i}}{2})=\lim_{\epsilon \to 0} e^{-\frac{2}{\epsilon}F}$
is a step function taking the values 0 and 1. We find (see appendix~C
and \cite{Fredenhagen:2004cj})
\begin{equation}\label{defofP}
P (\tfrac{p_{i}}{2}) =
\left\{ \begin{array}{ll}
1 & \lfloor p_{1}\rfloor +\lfloor p_{2}\rfloor +\lfloor p_{3}\rfloor
\ \text{even}\\
  & \text{and} \ |\{p_{1} \}-\{p_{2} \}|\leq \{p_{3} \} \leq \min
(\{p_{1}\} +\{p_{2}  \},2-\{p_{1} \}-\{p_{2} \})\\[1mm]
1& \lfloor p_{1}\rfloor +\lfloor p_{2}\rfloor +\lfloor p_{3}\rfloor
\ \text{odd}\\
  & \text{and} \ |\{p_{1} \}-\{p_{2} \}|\leq 1-\{p_{3} \} \leq \min
(\{p_{1}\} +\{p_{2}  \},2-\{p_{1} \}-\{p_{2} \})\\[1mm]
0 & \text{otherwise.}
\end{array} \right.
\end{equation}  
Now it remains to determine the phase $e^{i\eta (p_{i})}$. For the
values of the $p_{i}$ such that $F (\frac{p_{i}}{2})=0$, the phase
simplifies to
\begin{equation}\label{phasefactors}
e^{i\eta (p_{i})} = e^{-i\pi (\lfloor p_{1} \rfloor +\lfloor p_{2}
\rfloor + \lfloor p_{3} \rfloor ) 
+\frac{i\pi}{2}} \ .
\end{equation}
From the expressions~\eqref{defofP} and~\eqref{phasefactors} we see
that the three-point function is not analytic in the momenta any
more. To compare to the results from the limit of minimal models, we
have to rescale the fields by
\begin{equation}\label{rescaledfields}
V^{\text{NS}}_{\alpha} \to v^{\text{NS}}_{p}= i \frac{\epsilon}{2\pi}
\big( \mu \pi \gamma (\tfrac{bQ}{2})\big)^{\frac{1}{b} (\alpha -\frac{Q}{2})}
e^{i\pi \lfloor p \rfloor } V^{\text{NS}}_{\alpha} \ ,
\end{equation}
and the correlators on the sphere by
\begin{equation}\label{rescalingspherecor}
\langle \dotsb \rangle \to \frac{2\pi}{\epsilon^{2}} \langle \dotsb
\rangle \ .
\end{equation}
It is easy to see that with these rescalings and identifying
$p_{i}=\frac{d_{i}}{2}$, the limit $b\to i$ reproduces the same two-\
and three-point function
(see~\eqref{twopointfunction},\eqref{threepoint}) as the limit of
minimal models.  
\medskip

Let us now consider correlation functions involving a
superdescendant field
$\tilde{V}^{\text{NS}}_{\alpha}$. From~\cite{Fukuda:2002bv} we find
\begin{equation}
\langle V^{\text{NS}}_{\alpha_{1}} (z_{1})V^{\text{NS}}_{\alpha_{2}}
(z_{2})\tilde{V}^{\text{NS}}_{\alpha_{3}} (z_{3}) \rangle =
\tilde{C}^{\text{NS}}
(\alpha_{i}) |z_{12}|^{2 (h_{1}+h_{2}-\tilde{h}_{3})} |z_{23}|^{2
(h_{2}+\tilde{h}_{3}-h_{1})} |z_{13}|^{2 (h_{1}+\tilde{h}_{3}-h_{2})}
\ ,
\end{equation}
with
\begin{equation}
\tilde{C}^{\text{NS}} (\alpha_{i}) = i \big( \mu \pi \gamma
(\tfrac{bQ}{2})b^{1-b^{2}}\big)^{\frac{Q-2\tilde{\alpha}}{b}}\, 
\frac{2\Uns' (0) \Uns (2\alpha_{1}) \Uns (2\alpha_{2})\Uns
(2\alpha_{3})}{\Ur (2\tilde{\alpha}-Q)\Ur (2\tilde{\alpha}_{1})\Ur
(2\tilde{\alpha}_{2})\Ur (2\tilde{\alpha}_{3})} \ .
\end{equation}
The special functions $\Ur$ have been introduced in~\eqref{defofUnsUr}.
The superdescendant fields are normalised such that
\begin{equation}
\langle \tilde{V}^{\text{NS}}_{\alpha_{1}}
(z_{1})\tilde{V}^{\text{NS}}_{\alpha_{2}} (z_{2})\rangle =-4
h_{\alpha_{1}}^{2} |z_{12}|^{-2} \langle V_{\alpha_{1}} (z_{1})
V_{\alpha_{2}} (z_{2})\rangle \ .
\end{equation}
To perform the limit $b\to i$, we use the asymptotics of
the functions $\Uns$ and $\Ur$ given in~\eqref{asympU}. 
Inserting them into the expression for the correlator, we get
\begin{align}
\tilde{C}^{\text{NS}} (\alpha_{i}) & = 2 i \big( \mu \pi \gamma
(\tfrac{bQ}{2})\big)^{\frac{Q-2\tilde{\alpha}}{b}}
\frac{2\pi^{2}}{\epsilon} e^{-\frac{2}{\epsilon}F (\frac{p_{i}+1}{2})}
e^{i\eta  (p_{i}+1)} \nonumber\\
& \quad\ \times  \frac{\Upsilon (\frac{1}{2}-\tilde{p}|1)\Upsilon
(\frac{3}{2}-\tilde{p}|1) \prod_{i=1}^{3} \Upsilon
(\frac{1}{2}-\tilde{p}_{i}|1)\Upsilon
(\frac{3}{2}-\tilde{p}_{i}|1)}{\big[ \Upsilon (1|1)\Upsilon
(1-p_{1}|1)\Upsilon (1-p_{2}|1) \Upsilon (1-p_{3}|1)\big]^{2}} (1+o
(\epsilon^{0})) \ .
\end{align}
Rescaling 
\begin{equation}
\tilde{V}^{\text{NS}}_{\alpha} \to \tilde{v}^{\text{NS}}_{p}=-
\frac{1}{2h_{\alpha}} \frac{\epsilon}{2\pi} \big( \mu \pi \gamma
(\tfrac{bQ}{2})\big)^{\frac{1}{b} (\alpha -\frac{Q}{2})} e^{i\pi \lfloor
p\rfloor } \tilde{V}^{\text{NS}}_{\alpha} \ ,
\end{equation}
we recover in the limit $b\to i$ the corresponding three-point
correlator~\eqref{threepointdesc} that was obtained
from the limit of minimal models.

\subsection{Two- and three-point functions: Ramond sector}

In the normalisation of~\cite{Fukuda:2002bv} the two-point function of
two Ramond-Ramond fields is given by
\begin{align}
\langle \Theta_{\alpha_{1}}^{\pm \pm} (z_{1}) \Theta_{\alpha_{2}}^{\pm
\pm} (z_{2})\rangle & = |z_{12}|^{-4 h_{\alpha_{1}}-\frac{1}{4}}\, 2\pi
\delta (p_{1}+p_{2}) \\
\langle \Theta_{\alpha_{1}}^{\pm \pm} (z_{1}) \Theta_{\alpha_{2}}^{\mp
\mp} (z_{2}) \rangle & = |z_{12}|^{-4 h_{\alpha_{1}}-\frac{1}{4}}\, 2\pi
\delta (p_{1}-p_{2}) \tilde{R} (\alpha_{1}) \ ,
\end{align}
with the reflection amplitude
\begin{equation}
R^{\text{R}} (\alpha) 
=\big( \mu \pi \gamma (\tfrac{bQ}{2})\big)^{\frac{Q-2\alpha
}{b}} \frac{\Gamma (\frac{1}{2}+ipb) \Gamma (\frac{1}{2}+ip/b)}{\Gamma
(\frac{1}{2}-ipb) \Gamma (\frac{1}{2}-ip/b)} \ . 
\end{equation}
For the fields $V^{\text{R}}_{\alpha}$ (see~\eqref{newRRfields}) we
thus have the two-point function
\begin{equation}
\langle V^{\text{R}}_{\alpha_{1}} (z_{1}) V^{\text{R}}_{\alpha_{2}}
(z_{2})\rangle = |z_{12}|^{-4h_{\alpha_{1}}-\frac{1}{4}} \, 2\pi
\big(\delta (p_{1}+p_{2}) + \delta (p_{1}-p_{2}) R^{\text{R}}
(\alpha_{1}) \big) \ .
\end{equation}
The three-point functions involving RR fields are~\cite{Fukuda:2002bv}
(omitting the obvious coordinate dependence)
\begin{equation}
\langle V^{\text{NS}}_{\alpha_{1}} \Theta^{\pm \pm}_{\alpha_{2}} \Theta^{\mp
\mp}_{\alpha_{3}}\rangle = C^{\text{R}} (\alpha_{1},\alpha_{2},\alpha
_{3}) \quad ,\quad \langle V^{\text{NS}}_{\alpha_{1}} \Theta^{\pm
\pm}_{\alpha_{2}} \Theta^{\pm \pm}_{\alpha_{3}}\rangle =
\tilde{C}^{\text{R}} (\alpha_{1},\alpha_{2},\alpha_{3})\ , 
\end{equation}
with
\begin{align}
C^{\text{R}}(\alpha_i) & = \big( \mu\pi\gamma
(\tfrac{bQ}{2})b^{1-b^2}\big)^{\frac{Q-2\tilde{\alpha}}{b}}\,
\frac{\Upsilon_{\text{NS}}' (0)\Upsilon
_{\text{NS}}(2\alpha_1)\Upsilon _{\text{R}}(2\alpha_2)\Upsilon
_{\text{R}}(2\alpha_3)}{\Upsilon_{\text{R}}(2\tilde\alpha - Q)\Upsilon
_{\text{R}}(2\tilde\alpha_1)\Upsilon
_{\text{NS}}(2\tilde\alpha_2)\Upsilon
_{\text{NS}}(2\tilde\alpha_3)}  \\
\tilde{C}^{\text{R}}(\alpha_i) & = \big( \mu\pi\gamma
(\tfrac{bQ}{2})b^{1-b^2}\big)^{\frac{Q-2\tilde{\alpha}}{b}}
\frac{\Upsilon_{\text{NS}}' (0)\Upsilon
_{\text{NS}}(2\alpha_1)\Upsilon _{\text{R}}(2\alpha_2)\Upsilon
_{\text{R}}(2\alpha_3)}{\Upsilon_{\text{NS}}(2\tilde\alpha -
Q)\Upsilon _{\text{NS}}(2\tilde\alpha_1)\Upsilon
_{\text{R}}(2\tilde\alpha_2)\Upsilon _{\text{R}}(2\tilde\alpha_3)} \ .
\end{align}
The three-point function involving the fields $V^{\text{R}}_{\alpha}$
are then given by
\begin{equation}
\langle V^{\text{NS}}_{\alpha_{1}} V^{\text{R}}_{\alpha_{2}}
V^{\text{R}}_{\alpha_{3}}\rangle = C^{\text{R}} (\alpha_{i}) +
\tilde{C}^{\text{R}} (\alpha_{i}) \ .
\end{equation}
Let us now analyse the limit $b\to i$. The necessary formulae for the
asymptotic behaviour of $\Uns$ and $\Ur$ are given
in~\eqref{asympU}. In the limit, the structure constants
$C^{\text{R}}$ and $\tilde{C}^{\text{R}}$ behave as
\begin{align}
C^{\text{R}} (\alpha_{i}) & \sim \big( \mu \pi \gamma
(\tfrac{bQ}{2})\big)^{\frac{Q-2\tilde{\alpha}}{b}}
\frac{2\pi^{2}}{\epsilon} e^{-\frac{2}{\epsilon}F \big(
\frac{p_{1}}{2},\frac{p_{2}+\frac{1}{2}}{2},\frac{p_{3}+\frac{1}{2}}{2}\big)}
e^{i\eta (p_{1},p_{2}+\frac{1}{2},p_{3}+\frac{1}{2})+i\pi}\nonumber\\
 & \quad\ \times  \frac{\Upsilon (\frac{1}{2}-\tilde{p}|1) \Upsilon
(\frac{3}{2}-\tilde{p}|1) \Upsilon (\frac{1}{2}-\tilde{p}_{1}|1)
\Upsilon (\frac{3}{2}-\tilde{p}_{1}|1) 
\Upsilon (1-\tilde{p}_{2}|1)^{2} 
\Upsilon (1-\tilde{p}_{3}|1)^{2}}{\Upsilon (1|1)^{2} \Upsilon
(1-p_{1}|1)^{2} \Upsilon (\frac{1}{2}-p_{2}|1) \Upsilon
(\frac{3}{2}-p_{2}|1) \Upsilon (\frac{1}{2}-p_{3}|1) \Upsilon
(\frac{3}{2}-p_{3}|1)}  \\
\tilde{C}^{\text{R}} (\alpha_{i}) & \sim \big( \mu \pi \gamma
(\tfrac{bQ}{2})\big)^{\frac{Q-2\tilde{\alpha}}{2}}
\frac{2\pi^{2}}{\epsilon} e^{-\frac{2}{\epsilon}F \big( 
\frac{p_{1}}{2},\frac{p_{2}-\frac{1}{2}}{2},\frac{p_{3}+\frac{1}{2}}{2}\big)}
e^{i\eta (p_{1},p_{2}-\frac{1}{2},p_{3}+\frac{1}{2})}\nonumber\\
 & \quad\ \times \frac{\Upsilon (1-\tilde{p}|1)^{2} \Upsilon
(1-\tilde{p}_{1}|1)^{2} \Upsilon (\frac{1}{2}-\tilde{p}_{2}|1)
\Upsilon (\frac{3}{2}-\tilde{p}_{2}|1) \Upsilon
(\frac{1}{2}-\tilde{p}_{3}|1) \Upsilon
(\frac{3}{2}-\tilde{p}_{3}|1)}{\Upsilon (1|1)^{2} \Upsilon
(1-p_{1}|1)^{2} \Upsilon (\frac{1}{2}-p_{2}|1) \Upsilon
(\frac{3}{2}-p_{2}|1) \Upsilon (\frac{1}{2}-p_{3}|1) \Upsilon
(\frac{3}{2}-p_{3}|1)} \ .
\end{align}
Again we have to rescale the fields to obtain a finite limit. Choosing
the rescaling
\begin{equation}
\Theta_{\alpha}^{\pm \pm} \to \theta_{p}^{\pm \pm} =
\frac{\epsilon}{2\pi} \big( \mu \pi \gamma (\tfrac{bQ}{2})\big)^{ip/b}
e^{i\pi \lfloor p-\frac{1}{2}\rfloor} \Theta_{\alpha}^{\pm \pm}\ ,
\end{equation}
and similarly $V^{\text{R}}_{\alpha}\to v^{\text{R}}_{p} $, we obtain
for $b\to i$ the normalised two-point function
\begin{equation}
\langle v^{\text{R}}_{p_{1}} (z_{1}) v^{\text{R}}_{p_{2}}
(z_{2})\rangle = |z_{12}|^{-2p_{1}^{2}-\frac{1}{4}} \,
\big(\delta (p_{1}+p_{2}) + \delta (p_{1}-p_{2}) \big) \ .
\end{equation}
The limit of the three-point functions is then given by
\begin{align}
\langle v^{\text{NS}}_{p_{1}} \theta^{\pm \pm}_{p_{2}} \theta^{\mp
\mp}_{p_{3}}\rangle & = \frac{1}{2} P
(\tfrac{p_{1}}{2},\tfrac{p_{2}+\frac{1}{2}}{2},\tfrac{p_{3}+\frac{1}{2}}{2})
\nonumber\\
 & \quad\ \times  \frac{\Upsilon (\frac{1}{2}-\tilde{p}|1) \Upsilon
(\frac{3}{2}-\tilde{p}|1) \Upsilon (\frac{1}{2}-\tilde{p}_{1}|1)
\Upsilon (\frac{3}{2}-\tilde{p}_{1}|1) 
\Upsilon (1-\tilde{p}_{2}|1)^{2} 
\Upsilon (1-\tilde{p}_{3}|1)^{2}}{\Upsilon (1|1)^{2} \Upsilon
(1-p_{1}|1)^{2} \Upsilon (\frac{1}{2}-p_{2}|1) \Upsilon
(\frac{3}{2}-p_{2}|1) \Upsilon (\frac{1}{2}-p_{3}|1) \Upsilon
(\frac{3}{2}-p_{3}|1)} 
\label{CRlimit} \\
\langle v^{\text{NS}}_{p_{1}} \theta^{\pm \pm}_{p_{2}} \theta^{\pm
\pm}_{p_{3}}\rangle & = \frac{1}{2} P
(\tfrac{p_{1}}{2},\tfrac{p_{2}-\frac{1}{2}}{2},\tfrac{p_{3}+\frac{1}{2}}{2})
\nonumber\\
 & \quad\ \times \frac{\Upsilon (1-\tilde{p}|1)^{2} \Upsilon
(1-\tilde{p}_{1}|1)^{2} \Upsilon (\frac{1}{2}-\tilde{p}_{2}|1)
\Upsilon (\frac{3}{2}-\tilde{p}_{2}|1) \Upsilon
(\frac{1}{2}-\tilde{p}_{3}|1) \Upsilon
(\frac{3}{2}-\tilde{p}_{3}|1)}{\Upsilon (1|1)^{2} \Upsilon
(1-p_{1}|1)^{2} \Upsilon (\frac{1}{2}-p_{2}|1) \Upsilon
(\frac{3}{2}-p_{2}|1) \Upsilon (\frac{1}{2}-p_{3}|1) \Upsilon
(\frac{3}{2}-p_{3}|1)} \ .
\label{CRtildelimit}
\end{align}
Note that the correlators~\eqref{CRlimit} and~\eqref{CRtildelimit}
transform into each other when $p_{3}\to -p_{3}$.

\subsection{One-point functions on the upper half plane}

Boundary conditions for supersymmetric Liouville theory have been
analysed in~\cite{Ahn:2002ev,Fukuda:2002bv}. There is a discrete
family (the analogue of the ZZ-branes~\cite{Zamolodchikov:2001ah} in
bosonic Liouville theory) and a continuous family (corresponding to
the FZZT-branes~\cite{Fateev:2000ik,Teschner:2000md}). We shall
concentrate on the discrete family here and try to reproduce the
results that we obtained from the minimal models. These boundary conditions
are labelled by two integers $(u,v)$ where the sign of the gluing
condition for the supercurrent depends on $u+v$ being even or
odd. Similarly to the analysis for the minimal models we focus on the
case $u+v$ even. Then there are two boundary conditions $(u,v)_{\pm}$
for each tuple, differing in the sign of one-point functions for
RR-fields. 

The one-point functions (omitting the $z$-dependence) are~\cite{Fukuda:2002bv}
\begin{equation}
\langle V^{\text{NS}}_{\alpha} \rangle_{(u,v)_{\pm}} = -2\sqrt{\tfrac{2}{\pi}}
\big( \mu \pi \gamma (\tfrac{bQ}{2})\big)^{-ip/b} 
\, ip \, \Gamma (ipb) \Gamma (ip/b) \sinh \pi pub \, \sinh \pi pv/b
\end{equation}
and 
\begin{equation}
\langle V^{\text{R}}_{\alpha }\rangle_{(u,v)_{\pm}} = 
\pm 2^{3/4} \sqrt{\tfrac{2}{\pi}} \big( \mu \pi \gamma
(\tfrac{bQ}{2})\big)^{-\frac{ip}{b}} \, \Gamma (\tfrac{1}{2}+ipb) \Gamma
(\tfrac{1}{2}+\tfrac{ip}{b}) \sinh  (\pi pub + \tfrac{i\pi uv}{2})
\sinh (\tfrac{\pi pv}{b} - \tfrac{i\pi uv}{2}) \ .
\end{equation}
When we take the limit $b\to i$ we have to rescale the sphere
correlators by the factor $\frac{2\pi}{\epsilon^{2}}$
(see~\eqref{rescalingspherecor}). According to the discussion in
section~2.6 the scaling factor for the disc correlators should be
the square-root of it. To match the results of the minimal models we
have to choose the root with negative sign, so
\begin{equation}
\langle \dotsb \rangle_{(u,v)_{\pm}} \to -\frac{\sqrt{2\pi}}{\epsilon}
\langle \dotsb \rangle_{(u,v)_{\pm}} \ . 
\end{equation}
By going to the rescaled NSNS bulk fields $v^{\text{NS}}_{p}$
\eqref{rescaledfields} we find
\begin{align}
\langle v^{\text{NS}}_{p}\rangle_{(u,v)_{\pm}} & = \frac{1}{\pi}
e^{i\pi \lfloor p\rfloor} \Gamma (1-p)\Gamma (p) \sin \pi pu \sin \pi
pv\nonumber\\ 
& = \frac{\sin \pi pu \sin \pi pv}{|\sin \pi p|} \ ,
\end{align}
which coincides with the expression~\eqref{NSone-point-limit} we found
from the minimal models. For the RR fields we obtain similarly
\begin{equation}
\langle v^{\text{R}}_{p}\rangle_{(u,v)_{\pm}} = \pm
2^{\frac{3}{4}}\frac{\sin (\pi pu +\frac{\pi uv}{2}) \sin (\pi
pv+\frac{\pi u v}{2})}{|\cos \pi p|} \ ,
\end{equation}
and we find again complete agreement with the minimal model
result~\eqref{RRone-point-limit}.

\section{Summary and outlook} 

In this paper we have analysed the limit of supersymmetric minimal
models at central charge $c=\frac{3}{2}$. We computed the three-point
correlation functions for the limiting theory and showed that it is
possible to get the same correlators from a limit of supersymmetric
Liouville theory. In addition we obtained the one-point functions in
the presence of a boundary for a discrete family of boundary
conditions.

The computation of the three-point correlators cannot be seen as a
rigorous construction of the theory. The two ways of taking limits
both are singular: the transition from a discrete to a continuous
spectrum in the minimal model limit, and the loss of analyticity in
the limit of the Liouville correlators. Although the fact that we can
obtain the same correlators on two different paths makes it plausible
that the resulting theory is a healthy conformal field theory, this is
not guaranteed. Further checks on the theory should be performed, in
particular one should analyse whether the theory is crossing
symmetric. For the bosonic counterpart of the theory, this has been
analysed in~\cite{Runkel:2001ng}, on the one hand analytically for
specific four-point correlators, and on the other hand numerically for
the generic case. Similar checks could also be performed in the
supersymmetric case, numerical checks could use the recently found
recursion relations for superconformal
blocks~\cite{Hadasz:2006qb,Belavin:2006zr}. Instead of directly
testing the crossing symmetry, one would expect that it can be derived
from the crossing symmetry valid for the Liouville four-point correlators.

In this work we concentrated mostly on the Neveu-Schwarz sector of the
theory. In super Liouville theory, three-point functions involving
Ramond-Ramond fields have been determined and we used this result to
obtain their limit at $c=\frac{3}{2}$. On the other
hand, in the case of super minimal models the corresponding structure
constants have not been determined explicitly (see
however~\cite{Mussardo:1987eq,Mussardo:1987ab} for some results), so
before comparing the two limits it would be necessary to compute theses
quantities first.

We discussed a discrete family of boundary conditions and computed the
bulk one-point function. In addition there is another discrete family
(to which the RR fields do not couple) and also a continuous
family. It should be straightforward to determine the limits of
one-point functions for those. Further data, like bulk-boundary
operator expansion and boundary operator product expansion, however,
will be much more difficult to determine (for the bosonic case, the
limit of the bulk-boundary operator expansion for the continuous
family of branes was determined in~\cite{Fredenhagen:2004cj}, the
limit of the boundary OPE is still unknown).  
\medskip

One might now try to perform a similar analysis for the $N=2$ minimal
models and the $N=2$ Liouville theory. The $N=2$ minimal models can be
built as supersymmetric cosets $su (2)/u (1)$, and their structure
constants are
known~\cite{Zamolodchikov:1986bd,Christe:1986cy,Mussardo:1988av}.
They differ significantly from the bosonic or $N=1$ case, but it is
still possible to find analytic expressions that interpolate the
three-point functions to continuous $su (2)$
spins~\cite{Dabholkar:2007ey}, and it is conceivable that there exist
a non-trivial limit at $c=3$.\footnote{Aspects of the asymptotic
behaviour of the $su (2)$ structure constants have been analysed in a
different context in~\cite{D'Appollonio:2003dr}.} On the other hand,
also the Liouville structure constants are known in the $N=2$
case~\cite{Hosomichi:2004ph}, by a coset
construction~\cite{Giveon:1999px} they are related to those of the
$H_{3}^{+}$ model~\cite{Teschner:1997ft}. They are again built from
Barnes' double gamma function, but the limit that leads to central
charge $c=3$ now corresponds to taking $b\to 0$ instead of $b\to
i$. Again, a non-trivial limit might exist, but the asymptotics will
be completely different from the bosonic or $N=1$ case.

The limit of Virasoro minimal models can be viewed as a limit of
$su(2)$ coset models; analogously it should be possible to take such a
limit for all diagonal coset models of the type
\begin{equation*}
\frac{\mathfrak{g}_{k}\times  \mathfrak{g}_{1}}{\mathfrak{g}_{k+1}} \ ,
\end{equation*}
with some simply laced Lie algebra $\mathfrak{g}$. The limiting
central charge is $c=c (\mathfrak{g}_{1})=\text{rank}\,
\mathfrak{g}$. On the other hand, one can study the Toda conformal
field theory with Lie algebra $\mathfrak{g}$ whose central charge
approaches the same value $c=\text{rank}\, \mathfrak{g}$ in the limit
$b\to i$. One might suspect that the limits again lead to the same
conformal field theory, but as only a few explicit results on the
structure constants are available (see e.g.\ \cite{Fateev:2005gs} for
the conformal Toda theories and~\cite{Lukyanov:1990tf} for the minimal
models), we leave this analysis for future research. 

When we continue the expressions from bosonic Liouville theory to
$c=1$, one might wonder whether it is also possible to continue them
to $c<1$. This is in fact straightforward for rational central charges
of the form $c=1-6\frac{(p-q)^{2}}{pq}$. The corresponding conformal
field theories would not be unitary and would have a continuous
spectrum. Again the limit of the structure constants is the product of
an analytic part which coincides with the analytic interpolation of
the minimal model structure constants and a step function. Recently
this limit was investigated in~\cite{McElgin:2007ak}, and it was argued
that only for $p=1$ a physically sensible three-point function is
obtained. The relation of the resulting theory to minimal models or
generalised minimal models~\cite{Zamolodchikov:2005sj,Belavin:2006ex}
is unclear until now. The continuation of Liouville theory to other,
non-rational central charges $c<1$ is more singular, and it is clear
that the structure constants do not have a well-defined limit as
functions of the field parameters, though it might be that one can
find a sensible limit in terms of distributions.

\subsubsection*{Acknowledgements} 
We are grateful to Giuseppe D'Appollonio, Matthias Gaberdiel, Ingo
Runkel and Stefan Theisen for useful discussions. This paper is in
part based on the diploma thesis of one of the
authors~\cite{wellig:2006}. SF acknowledges support from the Swiss
National Science Foundation (SNF) for his time at ETH Z{\"u}rich where
the project was started.  

\appendix
\section{Structure constants in terms of Barnes' double
gamma functions}  
\subsection{Virasoro minimal models}

The unitary minimal models are labelled by an integer $p$; the
central charge is given by $c=1-\frac{6}{p (p+1)}$. The set of primary
fields $\phi_{rs}$ is specified by the Kac table, which consists of
pairs of integers $(r,s)$ with $1\leq r\leq p-1$ and $1\leq s\leq
p$. The conformal weight of a primary field $\phi_{rs}$ is given by
\begin{equation}
h_{rs} = \frac{((p+1)r-ps)^{2}-1}{4p(p+1)} \ .
\end{equation}
The pairs $(r,s)$ and $(p-r,p+1-s)$ label the same field. 

The operator product expansion for two primary fields is given by
\begin{equation}
\phi_{r_{1}s_{1}} (z_{1}) \phi_{r_{2}s_{2}} (z_{2}) \sim
\sum_{r_{3},s_{3}} \mc{N}^{(p)}_{r_{1}r_{2}}{}^{r_{3}}
\mc{N}^{(p+1)}_{s_{1}s_{2}}{}^{s_{3}}\, \,
\frac{D_{(r_{1}s_{1})(r_{2}s_{2})}{}^{(r_{3}s_{3})}}{|z_{12}|^{2(h_{r_{1}s_{1}}
+h_{r_{2}s_{2}}-h_{r_{3}s_{3}})}}
\phi_{r_{3}s_{3}} (z_{2}) \ . 
\end{equation}
Here, $\mc{N}^{(p)}\mc{N}^{(p+1)}$ implements the fusion constraints, and
$\mc{N}^{(p)}$ is defined as
\begin{equation}
\label{fusion}
\mc{N}^{(p)}_{r_{1}r_{2}}{}^{r_{3}} = \left\{\begin{array}{ll}
1 & |r_{1}-r_{2}|+1\leq r_{3} \leq  \min (r_{1}+r_{2}-1,2p-r_{1}-r_{2}-1),\
\ \sum_{i}r_{i} \ \text{odd}\\[1mm]
0 & \text{otherwise} \ .
\end{array} \right. 
\end{equation}
The structure constants $D$ have been determined originally by
Dotsenko and
Fateev~\cite{Dotsenko:1984nm,Dotsenko:1984ad,Dotsenko:1985hi}. We want
them to be normalised such that $D_{(rs) (rs)}{}^{(11)}=1$. It is
convenient to express them by structure constants $C$ in a different
normalisation of the fields,
\begin{align}
C_{(r_{1}s_{1}) (r_{2}s_{2})}{}^{(r_{3}s_{3})} & = t^{4 (k-1) (l-1)}
\prod_{m=1}^{k-1} \prod_{n=1}^{l-1} \big(d_{mn} (d_{mn}-d_{1})
(d_{mn}-d_{2}) (d_{mn}+d_{3}) \big)^{-2} \nonumber\\
& \quad \ \times  \prod_{m=1}^{k-1} \frac{\Gamma (\frac{m}{t}) \Gamma
(\frac{m-d_{1}}{t}) \Gamma (\frac{m-d_{2}}{t}) \Gamma
(\frac{m+d_{3}}{t})}{\Gamma (1-\frac{m}{t}) \Gamma
(1-\frac{m-d_{1}}{t}) \Gamma (1-\frac{m-d_{2}}{t} \Gamma
(1-\frac{m+d_{3}}{t}))} \nonumber \\
& \quad \ \times \prod_{n=1}^{l-1} \frac{\Gamma (nt)\Gamma (nt+d_{1}) \Gamma
(nt+d_{2}) \Gamma (nt-d_{3})}{\Gamma (1-nt) \Gamma (1-nt-d_{1}) \Gamma
(1-nt-d_{2}) \Gamma (1-nt+d_{3})} \ .
\label{Virstrucconst}
\end{align}
Here $k=\frac{r_{1}+r_{2}-r_{3}+1}{2}$,
$l=\frac{s_{1}+s_{2}-s_{3}+1}{2}$, $d_{mn}=m-nt$,
$d_{i}=d_{r_{i}s_{i}}$, $t=\frac{p}{p+1}$, and the normalised
structure constants can be recovered as
\begin{equation}\label{normalisedsc}
D_{(r_{1}s_{1}) (r_{2}s_{2})}{}^{(r_{3}s_{3})} =
\sqrt{\frac{C_{(r_{3}s_{3}) (r_{3}s_{3})}{}^{(11)}}{C_{(r_{1}s_{1})
(r_{1}s_{1})}{}^{(11)} C_{(r_{2}s_{2}) (r_{2}s_{2})}{}^{(11)}}}
\, C_{(r_{1}s_{1}) (r_{2}s_{2})}{}^{(r_{3}s_{3})} \ .
\end{equation}
The expression for the structure constants is also valid for the
non-unitary minimal models where $t=\frac{p}{q}$ is any rational
number. All results that will be obtained in this appendix will only
involve this parameter $t$, and therefore apply to all minimal models.

Dotsenko made the observation that the structure constants only depend
on a linear combination $d_{r_{i}s_{i}}$ of the Kac labels
$(r_{i},s_{i})$, and that they are analytic in the variables
$d_{r_{i}s_{i}}$. From the expression~\eqref{Virstrucconst} this is
not obvious because the Kac labels enter the range of products via $k$
and $l$. Although the result seems to be well known, there is, to our
knowledge, no explicit proof of this statement in the literature. We
therefore felt it could be useful to give a detailed derivation of
this fact here. In the following we shall re-express the structure
constants in terms of special functions, such that its analyticity in
$d_{r_{i}s_{i}}$ is manifest. The expression we derive has been given
recently in~\cite{Zamolodchikov:2005sj} for the three-point function
in generalised minimal models without explicitly showing that it
reproduces the Dotsenko-Fateev structure constants.

Let us briefly describe the general strategy. Suppose you are given a
finite product of functions with equally spaced arguments,
$\prod_{m=1}^{k-1} f (m)$, and you would like to obtain an expression
that is analytic in the range $k$ of the product and that interpolates
between integer values of $k$. If one can find an analytic function
$F$ with the shift property $F (x+1)=f (x) F (x)$, one can rewrite the
product as a quotient,
\begin{equation}
\prod_{m=1}^{k-1} f (m) = \frac{F (k)}{F (1)} \ ,
\end{equation}
which provides a manifestly analytic expression in $k$. This
expression is of course not unique and depends on the choice of $F$. 

The first product in the expression~\eqref{Virstrucconst} of the
structure constants is a double product in the variables $m,n$
over numbers depending linearly on $m$ and $n$. A product over equally
spaced numbers can be replaced by a ratio of gamma functions in two ways,
\begin{equation}\label{gammaratio}
\prod_{m=1}^{k-1} (m+x) = \frac{\Gamma (x+k)}{\Gamma (x+1)} =
(-1)^{k-1} \frac{\Gamma (-x) }{\Gamma (-x-k+1)} \ .
\end{equation}
By this means we can transform the double product into a single product
of quotients of gamma functions. The numbers in the double
product come as squares, so we can use both ways in~\eqref{gammaratio}
to rewrite the double product as
\begin{equation}
\prod_{m=1}^{k-1} \prod_{n=1}^{l-1} (x+m-nt)^{-2} = (-1)^{(k-1)
(l-1)}\prod_{n=1}^{l-1} \frac{\Gamma (x-nt+1)}{\Gamma (x-nt+k)} 
\frac{\Gamma (-x+nt-k+1)}{\Gamma (-x+nt)} \ ,
\end{equation}
where $x$ stands for the possible values $x=0,-d_{1},-d_{2},d_{3}$. 
Using this result we can write the structure constants $C$ as
\begin{equation}\label{Crewritten}
C_{(r_{1}s_{1}) (r_{2}s_{2})}{}^{(r_{3}s_{3})} = t^{4 (k-1) (l-1)}
\cdot \prod_{x=0,-d_{1},-d_{2},d_{3}} \!\! \Bigg( \prod_{m=1}^{k-1}
\frac{\Gamma (\frac{x+m}{t})}{\Gamma (1-\frac{x+m}{t})}
\prod_{n=1}^{l-1} \frac{\Gamma(-x+nt-k+1)}{\Gamma (x-nt+k)}\Bigg) \ .
\end{equation}
Now we only have to deal with products of gamma functions. 
The crucial step in rewriting the structure constants consists
in realising that a product over gamma functions with equally spaced
entries can be expressed as a quotient of Barnes' double gamma
functions,
\begin{equation}\label{gammatworatios}
\prod_{m=1}^{k-1} \Gamma (\tfrac{m+x}{t}) = (2\pi)^{\frac{k-1}{2}}\, 
t^{-\frac{k-1}{2t} (x+\frac{k}{2}-\frac{t}{2})}\, 
\frac{\Gamma_{2} \big( (1+x)t^{-\frac{1}{2}}\big| 
t^{-\frac{1}{2}}\big)}{\Gamma_{2}
\big( (k+x)t^{-\frac{1}{2}} \big| t^{-\frac{1}{2}}\big)} \ .
\end{equation}
This can be easily verified using the shift
property~\eqref{gammatwoshift} of the double gamma function
$\Gamma_{2}$.

With the help of identity~\eqref{gammatworatios} we replace the first
product of gamma functions in~\eqref{Virstrucconst} and obtain
\begin{equation}\label{upsilonratios}
\prod_{m=1}^{k-1} \frac{\Gamma
(\frac{m+x}{t})}{\Gamma (1-\frac{m+x}{t})} =
t^{-\frac{k-1}{t} (x+\frac{k}{2}-\frac{t}{2})}\, 
\frac{\Upsilon \big( (x+k)t^{-\frac{1}{2}}\big| t^{-\frac{1}{2}}\big)}{\Upsilon
\big( (x+1)t^{-\frac{1}{2}}\big| t^{-\frac{1}{2}}\big)} \ ,
\end{equation}
where the function $\Upsilon$ is defined by (see~\eqref{defofupsilon})
\begin{equation}
\Upsilon \big( y\big| t^{-\frac{1}{2}}\big) 
= \Big[ \Gamma_{2} \big( y\big| t^{-\frac{1}{2}}\big)
\Gamma_{2} \big( t^{\frac{1}{2}}+t^{-\frac{1}{2}}-y \big| 
t^{-\frac{1}{2}}\big) \Big]^{-1} \ .  
\end{equation}
We can apply the same procedure to the second product
in~\eqref{Crewritten}, bearing in mind that $\Gamma_{2} \big( x\big|
t^{-\frac{1}{2}}\big)=\Gamma_{2} \big( x\big| t^{\frac{1}{2}}\big)$. We
find
\begin{equation}
\prod_{n=1}^{l-1} \frac{\Gamma (-x+nt-k+1)}{\Gamma (x-nt+k)} =
t^{- (l-1)
(x+k-\frac{1}{2}-\frac{lt}{2})} \, \frac{\Upsilon \big(
(-x+lt-k+1)t^{-\frac{1}{2}}\big| t^{-\frac{1}{2}}\big)}{\Upsilon
\big( (x+k) t^{-\frac{1}{2}}\big| t^{-\frac{1}{2}}\big)}\ .
\end{equation}
Putting the results together we get
\begin{equation}
C_{(r_{1}s_{1}) (r_{2}s_{2})}{}^{(r_{3}s_{3})} = t^{- (1-t)
(\frac{k-1}{t}+ (l-1))} \cdot \prod_{x=0,-d_{1},-d_{2},d_{3}}
\frac{\Upsilon \big((-x-d_{kl}+1)t^{-\frac{1}{2}}
\big|t^{-\frac{1}{2}} \big)}{\Upsilon \big( (x+1)t^{-\frac{1}{2}}
\big| t^{-\frac{1}{2}} \big)} \ .
\end{equation}
Except for the exponent of $t$ in the factor in front, the structure
constant is expressed entirely in the variables $d_{i}$ (note that
$d_{kl}=\frac{d_{1}+d_{2}-d_{3}+1-t}{2}$). The $(r,s)$-dependent part
of the power of $t$ will drop out in the normalised structure
constants $D$. To determine them via~\eqref{normalisedsc} we
first observe that
\begin{equation}
C_{(rs) (rs)}{}^{(11)} = t^{- (1-t) (\frac{r-1}{t}+ (s-1))} \, 
\frac{\Upsilon \big( (d_{rs}+1)t^{-\frac{1}{2}}\big| t^{-\frac{1}{2}}
\big) \Upsilon \big( t^{-\frac{1}{2}} \big| t^{-\frac{1}{2}}
\big)}{\Upsilon \big( (d_{rs}+t)t^{-\frac{1}{2}} \big|
t^{-\frac{1}{2}} \big) \Upsilon
\big((2-t)t^{-\frac{1}{2}}\big|t^{-\frac{1}{2}}\big)} \ . 
\end{equation}
The normalised structure constants $D$ then only depend on the
variables $d_{i}$ and we obtain
\begin{equation}\label{strucconstresult}
D_{(r_{1}s_{1}) (r_{2}s_{2})}{}^{(r_{3}s_{3})} = D_{0} 
 \prod_{x=0,-d_{1},-d_{2},d_{3}} \!\!\frac{\Upsilon
\big((-x-d_{kl}+1)t^{-\frac{1}{2}}\big|t^{-\frac{1}{2}}
\big)}{\Big[ \Upsilon\big((x+1)t^{-\frac{1}{2}}\big| t^{-\frac{1}{2}}
\big) \Upsilon \big( (x+t)t^{-\frac{1}{2}}\big| t^{-\frac{1}{2}} \big)
\Big]^{\frac{1}{2}}} \ ,
\end{equation}
with the normalisation constant
\begin{equation}
D_{0}= t^{-\frac{t-t^{-1}-1}{2}} \big[ \gamma (1-t^{-1}) \gamma (2-t)
\big]^{-\frac{1}{2}} \ .
\end{equation}
Note that our derivation is also valid for the non-unitary minimal
models where $t=\frac{p}{q}$ is any rational number. The
expression~\eqref{strucconstresult} coincides with the formula given
in \cite{Zamolodchikov:2005sj}. The normalised structure constants $D$ are
invariant under permutation of the variables $d_{i}$. Furthermore,
they do not change if one performs field identifications in two field
labels (e.g.\ $d_{1}\to -d_{1}$ and $d_{2}\to -d_{2}$) at the same time. 

Let us discuss the uniqueness of our result. For a fixed
minimal model there are only a finite number of fields, and the
analytic interpolation of the structure constants is of course not
unique. As already stated above, the Dotsenko-Fateev expressions apply
to any rational $t$. If we require the interpolation to be continuous
in the parameter $t$, then it is uniquely fixed: any given $d$ and $t$
can be obtained as a limit of rational $d_{r_{i}s_{i}}$ and $t_{i}$.

\subsection{Superconformal minimal models}

The $N=1$ supersymmetric minimal models are labelled by an integer $p$,
and they have central charge 
\begin{equation}
c= \frac{3}{2}\bigg(1-\frac{8}{p (p+2)} \bigg)\ .
\end{equation}
The primary fields $\phi_{rs}$ are labelled by pairs of integers $(r,s)$ in the
range $1\leq r\leq p-1$ and $1\leq s\leq p+1$. Two pairs $(r,s)$ and
$(p-r,p+2-s)$ label the same field. Depending on the
sum $r+s$ being even or odd, the corresponding primary field belongs
to the Neveu-Schwarz or Ramond sector, respectively.

We shall only consider the Neveu-Schwarz sector. The operator product
expansion of two primary fields is given by
\begin{align}
\phi_{r_{1}s_{1}} (z_{1}) \phi_{r_{2}s_{2}} (z_{2}) 
& \sim \sum_{r_{3},s_{3}} 
\mc{N}^{(p)}_{r_{1}r_{2}}{}^{r_{3}}
\mc{N}^{(p+2)}_{s_{1}s_{2}}{}^{s_{3}}\nonumber\\
& \quad \ \times  \bigg( \frac{\delta_{2}
(k+l)}{|z_{12}|^{h_{1}+h_{2}-h_{3}}}
D^{\text{NS}}_{(r_{1}s_{1}) (r_{2}s_{2})}{}^{(r_{3}s_{3})}
\phi_{r_{3}s_{3}} (z_{2}) \nonumber\\
&\qquad\quad  +\frac{\delta_{2}
(k+l+1)}{|z_{12}|^{h_{1}+h_{2}-\tilde{h}_{3}}}
\tilde{D}^{\text{NS}}_{(r_{1}s_{1}) (r_{2}s_{2})}{}^{(r_{3}s_{3})}
\tilde{\phi}_{r_{3}s_{3}} (z_{2}) \bigg) \ ,
\label{N1OPE}
\end{align}
where $k=\frac{r_{1}+r_{2}-r_{3}+1}{2}$, $l=\frac{s_{1}+s_{2}-s_{3}+1}{2}$.
We used the notation $\delta_{2} (m)$ for a function that is $1$ if
$m$ is even, and $0$ if it is odd. The conformal weights of the fields
are denoted by $h_{i}=h^{\text{NS}}_{r_{i}s_{i}}$ with
\begin{equation}
h^{\text{NS}}_{rs} = \frac{\big((p+2)r-ps \big)^{2}-4}{8p (p+2)} \ .
\end{equation}
The field $\tilde{\phi}_{rs}$ denotes the superpartner to the field
$\phi_{rs}$ with conformal weight $\tilde{h}^{\text{NS}}_{rs}$.  

As in the case of Virasoro minimal models, it is more convenient to
state the structure constants $C,\tilde{C}$ for a different
normalisation of the fields. The normalised structure constants are
recovered again by~\eqref{normalisedsc}. These constants have been
determined in~\cite{Kitazawa:1987za} and can be written as
\begin{align}
C^{\text{NS}}_{(r_{1}s_{1}) (r_{2}s_{2})}{}^{(r_{3}s_{3})} & =
\prod_{x=0,-d_{1},-d_{2},d_{3}}
\Bigg(\prod_{n=1}^{l-1}\prod_{\substack{m=1\\
\makebox[0cm][c]{$\mspace{-26mu}\scriptstyle m+n\ \text{even}$}}}^{k-1}
\frac{t}{(\frac{d_{mn}+x}{2})^{2}}   \nonumber\\
& \quad\ \times 
\prod_{m=1}^{k-1} \frac{\Gamma (\frac{x+m+t\delta_{2} (m+1)}{2t})}{\Gamma
(1-\frac{x+m+t\delta_{2} (m+1)}{2t})} 
\prod_{n=1}^{l-1} \frac{\Gamma (\frac{nt-x+\delta_{2}
(n+1)}{2})}{\Gamma (1-\frac{nt-x+\delta_{2} (n+1)}{2})}\Bigg) \ ,
\label{sstrucconst}
\end{align}
and
\begin{equation}
\tilde{C}^{\text{NS}}_{(r_{1}s_{1}) (r_{2}s_{2})}{}^{(r_{3}s_{3})} = 
C^{\text{NS}}_{(r_{1}s_{1}) (r_{2}s_{2})}{}^{(r_{3}s_{3})} \cdot  \left\{
\begin{array}{ll} \frac{2}{\big(\frac{r_{3}+1}{2t}-\frac{s_{3}+1}{2}
\big)^{2}}\ , & \ k\ \text{even},\ l\ \text{odd}\\[1mm]
\frac{2}{t^{2}\big(\frac{r_{3}+1}{2t}-\frac{s_{3}+1}{2} \big)^{2}}  
\ , & \ k\ \text{odd},\ l\ \text{even} \ .
\end{array} \right.
\end{equation}
Here, $d_{mn}=m-nt$, $d_{i}=d_{r_{i}s_{i}}$ and $t=\frac{p}{p+2}$.

Let us now try to find an expression for the structure constants that
only depends on the variables $d_{i}$. The strategy is similar to the
one pursued in the case of Virasoro minimal models in the previous
section. Let us first rewrite the double product in the first line
of~\eqref{sstrucconst} as
\begin{align}
\prod_{n=1}^{l-1}\prod_{\substack{m=1\\
\makebox[0cm][c]{$\mspace{-26mu}\scriptstyle m+n\
\text{even}$}}}^{k-1} \frac{t}{\big(\frac{d_{mn}+x}{2}\big)^{2}} 
& = (-t)^{\big\lfloor\frac{k}{2}\big\rfloor 
\big\lfloor\frac{l}{2}\big\rfloor 
+ \big\lfloor\frac{k-1}{2}\big\rfloor
\big\lfloor\frac{l-1}{2}\big\rfloor} \nonumber\\
& \quad \ \times  
\prod_{n=1}^{l-1} \frac{\Gamma \big(1-\frac{nt-x+\delta_{2}
(n+1)}{2}\big)}{\Gamma \big(\frac{x-nt+k+1-\delta_{2} (k+n)}{2}\big)}
\frac{\Gamma \big(1-\frac{x-nt+k+1-\delta_{2} (k+n)}{2}\big)}{\Gamma
\big(\frac{nt-x+\delta_{2} (n+1)}{2}\big)}\ .
\end{align}
The structure constants then become
\begin{align}
C^{\text{NS}}_{(r_{1}s_{1}) (r_{2}s_{2})}{}^{(r_{3}s_{3})} & =
 (t^{4})^{\big\lfloor\frac{k}{2}\big\rfloor 
\big\lfloor\frac{l}{2}\big\rfloor 
+ \big\lfloor\frac{k-1}{2}\big\rfloor
\big\lfloor\frac{l-1}{2}\big\rfloor}
\cdot \prod_{x=0,-d_{1},-d_{2},d_{3}}\nonumber\\
&\quad \ \times 
\Bigg( \prod_{m=1}^{k-1} \frac{\Gamma \big(\frac{x+m+t\delta_{2}
(m+1)}{2t}\big)}{\Gamma \big(1-\frac{x+m+t\delta_{2} (m+1)}{2t}\big)}
\prod_{n=1}^{l-1} \frac{\Gamma \big(1-\frac{x-nt+k+1-\delta_{2}
(k+n)}{2}\big)}{\Gamma \big(\frac{x-nt+k+1-\delta_{2} (k+n)}{2}\big)}\Bigg) \ .
\end{align}
Now we replace the products of gamma functions by ratios of Barnes'
double gamma functions. Take the first product over $m$. We split it
up into two products, one over even $m$ and one over odd
$m$. Employing the formula~\eqref{upsilonratios} for the separate
products then yields
\begin{align}
\prod_{m=1}^{k-1}\frac{\Gamma \big(\frac{x+m+t\delta_{2}
(m+1)}{2t}\big)}{\Gamma \big(1-\frac{x+m+t\delta_{2} (m+1)}{2t} \big)}
& = t^{-\frac{1}{2t}\big\lfloor\frac{k}{2}\big\rfloor\big(x
+\big\lfloor\frac{k}{2} \big\rfloor \big)} \cdot t^{-\frac{1}{2t}
\big\lfloor \frac{k-1}{2}\big\rfloor
\big(x+1-t+\big\lfloor\frac{k-1}{2} \big\rfloor \big)}\nonumber\\
& \quad \ \times 
\frac{\Upsilon \big(\big(\frac{x+1+t}{2} +\big\lfloor\frac{k}{2}
\big\rfloor \big) t^{-\frac{1}{2}} \big| t^{-\frac{1}{2}}\big)
\Upsilon \big( \big(\frac{x+2}{2}+\big\lfloor \frac{k-1}{2}
\big\rfloor \big)t^{-\frac{1}{2}} \big| t^{-\frac{1}{2}}
\big)}{\Upsilon \big( \frac{x+1+t}{2}t^{-\frac{1}{2}}\big|
t^{-\frac{1}{2}} \big) \Upsilon \big( \frac{x+2}{2}t^{-\frac{1}{2}}
\big| t^{-\frac{1}{2}} \big)} \ .
\end{align}
Doing the same with the product over $n$ leads to
\begin{multline}
\prod_{n=1}^{l-1} \frac{\Gamma \big(1-\frac{x-nt+k+1-\delta_{2}
(k+n)}{2}  \big)}{\Gamma \big( \frac{x-nt+k+1-\delta_{2}
(k+n)}{2}\big)} =
t^{-\big\lfloor \frac{l}{2}\big\rfloor \big(\frac{x}{2} +\big\lfloor
\frac{k}{2}\big\rfloor -\frac{t}{2}\big\lfloor\frac{l}{2}
\big\rfloor\big)} \cdot t^{-\big\lfloor\frac{l-1}{2} \big\rfloor \big(
\frac{x-t+1}{2}+\big\lfloor\frac{k-1}{2}
\big\rfloor-\frac{t}{2}\big\lfloor\frac{l-1}{2} \big\rfloor\big)} \\
\times \frac{\Upsilon \big(\big(\frac{x+1+t}{2}+\big\lfloor\frac{k}{2}
\big\rfloor-t\big\lfloor\frac{l}{2} \big\rfloor\big)t^{-\frac{1}{2}}
\big| t^{-\frac{1}{2}} \big) 
\Upsilon \big(\big(\frac{x+2}{2} +\big\lfloor\frac{k-1}{2} \big\rfloor
-t\big\lfloor\frac{l-1}{2}
\big\rfloor\big)t^{-\frac{1}{2}}\big|t^{-\frac{1}{2}} \big)}{
\Upsilon\big(\big(\frac{x+1+t}{2} +\big\lfloor\frac{k}{2}
\big\rfloor\big)t^{-\frac{1}{2}} \big| t^{-\frac{1}{2}} \big)
\Upsilon \big(\big(\frac{x+2}{2} +\big\lfloor\frac{k-1}{2}
\big\rfloor\big) t^{-\frac{1}{2}}\big|t^{-\frac{1}{2}} \big)}\ .
\end{multline}
When we combine the results, we obtain
\begin{multline}
C^{\text{NS}}_{(r_{1}s_{1}) (r_{2}s_{2})}{}^{(r_{3}s_{3})} =
t^{- (1-t)\big(\frac{k-1}{2t} +\frac{l-1}{2} \big) +\delta_{2}
(l)-\delta_{2} (k)} \\
\times \!\!\!\!\prod_{x=0,-d_{1},-d_{2},d_{3}} 
 \frac{\Upsilon \big(\big(\frac{x+1+t+d_{kl}}{2}+\frac{t\delta_{2} (k+l+1)}{2}
 \big)t^{-\frac{1}{2}}
\big| t^{-\frac{1}{2}} \big) \Upsilon \big(\big(\frac{x+2t+d_{kl}}{2}
-\frac{t\delta_{2} (k+l+1)}{2}\big)t^{-\frac{1}{2}}
\big|t^{-\frac{1}{2}} \big)}{\Upsilon
\big( \frac{x+1+t}{2}t^{-\frac{1}{2}}\big| t^{-\frac{1}{2}} \big)
\Upsilon \big( \frac{x+2}{2}t^{-\frac{1}{2}} \big| t^{-\frac{1}{2}}
\big)} \ .
\end{multline}
Except for the exponent of $t$ we get a result that only
depends on the variables $d_{i}$ and on $k+l$ being even or odd,
corresponding to even or odd fusion.

To obtain the normalised structure constants $D^{\text{NS}}$ we
first determine
\begin{equation}
C^{\text{NS}}_{(rs) (rs)}{}^{(11)} = t^{-
(1-t)\big(\frac{r-1}{2t}+\frac{s-1}{2} \big)} 
\frac{\Upsilon \big(\frac{2+d_{rs}}{2}t^{-\frac{1}{2}}\big|
t^{-\frac{1}{2}} \big) \Upsilon
\big(\frac{1+t}{2}t^{-\frac{1}{2}}\big|t^{-\frac{1}{2}} \big)}{\Upsilon
\big(\frac{2t+d_{rs}}{2}t^{-\frac{1}{2}} \big| t^{-\frac{1}{2}} \big)
\Upsilon \big(\frac{3-t}{2}t^{-\frac{1}{2}} \big| t^{-\frac{1}{2}} \big)} 
\ .
\end{equation}
The normalised structure constants are then given by
\begin{equation}\label{mmstruc}
D^{\text{NS}}_{(r_{1}s_{1}) (r_{2}s_{2})}{}^{(r_{3}s_{3})} =
D_{0}^{\text{NS}} \!\! \prod_{x=0,-d_{1},-d_{2},d_{3}} \frac{\Upsilon
\big(\frac{x+1+t+d_{kl}}{2}t^{-\frac{1}{2}} \big| t^{-\frac{1}{2}}
\big) \Upsilon \big(\frac{x+2t+d_{kl}}{2}t^{-\frac{1}{2}} \big|
t^{-\frac{1}{2}} \big)}{\Upsilon \big(\frac{x+1+t}{2}t^{-\frac{1}{2}}
\big| t^{-\frac{1}{2}} \big) \big[\Upsilon\big(
\frac{x+2}{2}t^{-\frac{1}{2}}\big| t^{-\frac{1}{2}} \big) \Upsilon
\big(\frac{x+2t}{2}t^{-\frac{1}{2}}\big| t^{-\frac{1}{2}} \big)
\big]^{\frac{1}{2}}} \ ,
\end{equation}
with the normalisation constant
\begin{equation}
D^{\text{NS}}_{0} = t^{-\frac{1}{4} (t-t^{-1}+2)} \big[\gamma
(\tfrac{t+1}{2}) \gamma (\tfrac{t^{-1}-1}{2}) \big]^{\frac{1}{2}} \ .
\end{equation}
The structure constants~\eqref{mmstruc} are invariant under
permutation of the field labels, and do not change if one applies a
field identification to two of the field labels, e.g.\ $d_{1}\to
-d_{1}$ and $d_{2}\to -d_{2}$. The structure constants for the OPE
involving superdescendants are given by
\begin{equation}\label{mmstrucdesc}
\tilde{D}^{\text{NS}}_{(r_{1}s_{1}) (r_{2}s_{2})}{}^{(r_{3}s_{3})} =
\tfrac{8t\, D_{0}^{\text{NS}}}{(d_{3}+1-t)^{2}}
\prod_{\begin{subarray}{c}x=0,-d_{1},\\ -d_{2},d_{3}\end{subarray}}
\frac{\Upsilon
\big(\frac{x+1+2t+d_{kl}}{2}t^{-\frac{1}{2}} \big| t^{-\frac{1}{2}}
\big) \Upsilon \big(\frac{x+t+d_{kl}}{2}t^{-\frac{1}{2}} \big|
t^{-\frac{1}{2}} \big)}{\Upsilon \big(\frac{x+1+t}{2}t^{-\frac{1}{2}}
\big| t^{-\frac{1}{2}} \big) \big[\Upsilon\big(
\frac{x+2}{2}t^{-\frac{1}{2}}\big| t^{-\frac{1}{2}} \big) \Upsilon
\big(\frac{x+2t}{2}t^{-\frac{1}{2}}\big| t^{-\frac{1}{2}} \big)
\big]^{\frac{1}{2}}} \ .
\end{equation}
Similar to the discussion at the end of appendix~A.1, given the
structure constants for all superconformal minimal models with
rational $t$, the provided expressions~(\ref{mmstruc},\ref{mmstrucdesc})  
give the unique interpolation that is continuous both in $d$ and in $t$.

\section{Special functions}
Barnes' double gamma function $\Gamma_{2} (x|b,b^{-1})$ is defined
for $x\in \mathbb{C}$ and complex $b$ with $\text{Re}\, b \not= 0$ (see
\cite{Barnes1}), and can be written as
\begin{equation}
\log \Gamma_{2} (x|b,b^{-1}) = \bigg(\frac{\partial}{\partial t
}\sum_{n_{1},n_{2}=0}^{\infty } (x+n_{1}b+n_{2}b^{-1})^{-t}
\bigg)_{t=0} \ . 
\end{equation}
We shall work with a different normalisation,
$\Gamma_{2} (x|b):=\Gamma_{2}(x|b,b^{-1})/\Gamma_{2} (Q/2|b,b^{-1})$. 
The logarithm of $\Gamma_{2}$ can be represented by an integral,
\[
\log \Gamma_{2}(x|b) = -\int _{0}^{\infty } \frac{{\rm
d}t}{t} \Bigg( \frac{e^{-Qt/2}-e^{-xt}}{(1-e^{-bt})
(1-e^{-t/b})}+\frac{(Q/2-x)^{2}}{2}e^{-t}+\frac{Q/2-x}{t}  \Bigg) \ ,
\]
where
$Q=b+b^{-1}$. The integral converges for $x$ with $\text{Re}\, x >0$. 

The double gamma function satisfies the functional relations
\begin{align}
\Gamma_{2}(x+b|b) &= \frac{\sqrt{2\pi}}{\Gamma (bx)} 
b^{-1/2+bx}\Gamma_{2}(x|b)
\label{gammatwoshift}\\
\Gamma_{2}(x+1/b|b) &= \frac{\sqrt{2\pi}}{\Gamma (x/b)}
b^{-x/b+1/2}\Gamma_{2} (x|b) \ .
\end{align}
Further properties of the double gamma function can be found in the
literature (see e.g.\ \cite{Jimbo:1996ss}). We also need the combination
\begin{equation}\label{defofupsilon}
\Upsilon (x|b) = \Gamma_{2}^{-1} (x|b)\Gamma_{2}^{-1} (Q-x|b)\ .
\end{equation}
The function $\Upsilon$ also has a simple behaviour under shifts of
the argument by $b$ or $b^{-1}$,
\begin{align}
\Upsilon (x+b|b) =& \gamma (bx) b^{1-2bx}
\Upsilon (x|b)\\
\Upsilon (x+b^{-1}|b) =& \gamma (b^{-1}x)
b^{-1+2xb^{-1}} \Upsilon (x|b) \ ,
\end{align}
where $\gamma (x)=\frac{\Gamma (x)}{\Gamma (1-x)}$.

\section{Fusion rules from Liouville}

In the $b\to i$ limit of Liouville theory, the fusion rules of the
limiting theory at $c=\frac{3}{2}$ arise from the term
$e^{-\frac{2}{\epsilon}F (\frac{p_{i}}{2})}$ in the limit $\epsilon
\to 0$, with the function $F$ (see~\eqref{defofF}) given by
\begin{equation}
F\Big(\frac{p_i}{2}\Big) = \sum _{i=1}^{3} (\lambda (p_i) - \lambda
(\tilde p_i)) - \lambda (\tilde p) + \frac{\pi ^2}{2} \ .
\end{equation}
The function $\lambda$ is given after~\eqref{asympUpsilon}. Here we
want to prove that $F$ is always greater or equal to zero (so the
limit of $e^{-\frac{2}{\epsilon}F}$ is well defined), in particular
\begin{equation}
F\Big(\frac{p_i}{2}\Big) = \left\{\begin{array}{ll} 0 & \lfloor p_1
\rfloor + \lfloor p_2 \rfloor +\lfloor p_3 \rfloor \quad
\textrm{even}\\ 
& \textrm{and} \quad |\{p_1\} - \{p_2\} | \leq \{p_3\}\leq
\textrm{min} (\{p_1\}+\{p_2\} , 2- \{p_1\}-\{p_2\} )\\[1mm] 
0 & \lfloor p_1 \rfloor + \lfloor p_2 \rfloor +\lfloor p_3
\rfloor\quad  \textrm{odd}\\ 
& \textrm{and}\quad  |\{p_1\} - \{p_2\} | \leq 1- \{p_3\}\leq
\textrm{min} (\{p_1\}+\{p_2\} , 2- \{p_1\}-\{p_2\} )\\[1mm]
\text{positive} & \textrm{otherwise} .\label{condition} \end{array}\right.
\end{equation}
To study the function $F$ we have to relate the fractional part of
$\tilde{p}$ and $\tilde{p}_{i}$ to the fractional parts $\{p_{i}\}$.
We first look at the case when $\lfloor p_1\rfloor + \lfloor p_2
\rfloor +\lfloor p_3 \rfloor$ is even. We find
\begin{align}
\{\tilde p\} &= \bigg\{ \frac{\{p_1\}+\{p_2\}+\{p_3\}}{2}\bigg\}
=\left\{\begin{array}{ll} \frac{\{p_1\}+\{p_2\}+\{p_3\}}{2}, &\{p_3\}
< 2- \{p_1\}-\{p_2\}\\[1mm]
\frac{\{p_1\}+\{p_2\}+\{p_3\}}{2}-1, & \{p_3\} \ge 2- \{p_1\}-\{p_2\}
\end{array}\right. \\
\{\tilde p_i\} &= \bigg\{ \frac{\sum_{j}\{p_j\} -2\{p_{i} \}}{2}\bigg\}
=\left\{\begin{array}{ll} \frac{\sum_{j}\{p_j\}-2\{p_i\}}{2}, &
2\{p_i\} \leq \sum_{j}\{p_j\}  \\[1mm]
\frac{\sum_{j}\{p_j\}-2\{p_i\}}{2}+1, & 2\{p_i\} > \sum_{j}\{p_j\}
\end{array}\right.  \ .
\end{align}
Evaluating $\lambda (\tilde{p}),\lambda (\tilde{p}_{i})$ we get
\begin{align}
\label{lambdaptilde}
\lambda (\tilde p) & = \left\{ \begin{array}{ll}
\frac{\pi ^2}{2} (\sum_{j}\{p_j\}-1)^2 \ ,\ & \{p_3\}
\leq 2- \{p_1\}-\{p_2\}\\[1mm]
\frac{\pi^{2}}{2} (\sum_{j}\{p_{j} \}-1)^{2} -2\pi^{2} (\sum_{j}\{p_{j}\}-2)
\ ,\ & \{p_3\}> 2- \{p_1\}-\{p_2\} \end{array} \right.\\
\label{lambdaptildei}
\lambda (\tilde{p}_{i}) & = \left\{\begin{array}{ll}
\frac{\pi ^2}{2} (\sum_{j}\{p_j\}-2\{p_i\}-1)^2\ ,\  & 2\{p_i\} \leq
\sum_{j}\{p_j\}\\[1mm] 
\frac{\pi ^2}{2} (\sum_{j}\{p_j\}-2\{p_i\}-1)^2 -2\pi^{2}
(2\{p_{i} \}-\sum_{j}\{p_{j}\})\ ,\  & 2\{p_i\} >
\sum_{j}\{p_j\} \ .
\end{array} \right.
\end{align}
Let us consider the case when the first lines
of~\eqref{lambdaptilde} and~\eqref{lambdaptildei} apply. A
straightforward calculation shows that $F=0$. This corresponds to the
first case in~\eqref{condition}. It is easy to see that in any other
case, the function $F$ obtains an additional term that makes it
positive. This concludes the case $\sum_{j}\lfloor p_{j}\rfloor $
even.      

If $\sum_{j}\lfloor p_{j}\rfloor$ is odd, we can still use the
analysis above if we replace $p_i \rightarrow - p_i$. The function $F$
is invariant under this replacement, and for non-integer $p_{i}$ we
have $\lfloor -p_{i}\rfloor = -\lfloor p_{i}\rfloor -1$, so for
generic numbers $p_{i}$ the sum $\sum_{i}\lfloor -p_{i}\rfloor$ is
again even and we are back to the case we considered before. The
fractional parts are then related by $\{-p_{i} \}=1-\{p_{i}\}$, and in
this way we obtain the second line of the result for $F$
in~\eqref{condition} from the first line. If one or more labels
$p_{i}$ are integer, the same arguments apply: we can still replace
$\{p_{i}\}$ by $1-\{p_{i}\}$ (which maps $0$ to $1$) to relate the
problem to the old analysis (we there did not use the fact that the
fractional parts were strictly less than $1$).


\end{document}